\documentclass{jfm}
\usepackage{graphicx}
\usepackage{epstopdf, epsfig}
\usepackage{bm}
\usepackage{amssymb}
\usepackage{tasks}
\usepackage{amsbsy}
\usepackage{amsmath}
\usepackage{mathtools}
\usepackage{xfrac}
\usepackage{float}
\usepackage{graphicx,calc}
\usepackage{tikz}
\usepackage{mathrsfs}
\usepackage{caption}
\usepackage{subcaption}
\usepackage{enumitem,lipsum}
\usepackage{pstricks}

\usetikzlibrary{arrows.meta}
\usetikzlibrary{calc}
\usetikzlibrary{arrows,decorations.pathmorphing,backgrounds,intersections,positioning,fit,matrix}
\usetikzlibrary{arrows,calc,shapes,decorations.pathreplacing}
\tikzstyle{doublearr}=[latex-latex, black, line width=0.5pt]
\usepackage[font=small]{caption}
\usepackage[labelformat = empty,position=top]{subcaption}
\usepackage[export]{adjustbox}
\DeclareRobustCommand\sampleline[1]{%
  \tikz\draw[#1] (0,0) (0,\the\dimexpr\fontdimen22\textfont2\relax)
  -- (1.5em,\the\dimexpr\fontdimen22\textfont2\relax);%
}
\newcommand{\BEQ}{\begin{equation}}
\newcommand{\EEQ}{\end{equation}}
\newcommand{\GM}{G_{\text{max}}}
\newcommand{\TO}{\tau_{\text{opt}}}
\newcommand{\KO}{k_{\text{opt}}}
\newcommand{\BU}{\boldsymbol{u}}
\newcommand{\BN}{\boldsymbol{\nabla}}
\newcommand{\EY}{\boldsymbol{e}_y}
\newcommand{\thd}{three-dimensional~}
\newcommand\Ha{\mathit{Ha}}
\DeclareSymbolFont{newfont}{OML}{cmm}{m}{it}
\DeclareMathSymbol{\Epsilon}{0}{newfont}{15}

\newcommand*{\rom}[1]{\expandafter\@slowromancap\romannumeral #1@}
\newcommand{\appropto}{\mathrel{\vcenter{
  \offinterlineskip\halign{\hfil$##$\cr
    \propto\cr\noalign{\kern2pt}\sim\cr\noalign{\kern-2pt}}}}}

\newlength\myheight
\newlength\mydepth
\settototalheight\myheight{Xygp}
\shorttitle{From 3D to Q2D: Transient growth in MHD duct flows}
\shortauthor{O. G. W. Cassells, T. Vo, A. Poth\'erat and G. J. Sheard}

\title{From three-dimensional to quasi-two-dimensional: Transient growth in magnetohydrodynamic duct flows}

\author{Oliver G. W. Cassells\aff{1},
  Tony Vo\aff{1},
  Alban Poth\'erat\aff{2}
 \and Gregory J. Sheard\aff{1}
 \corresp{\email{greg.sheard@monash.edu}}}

\affiliation{\aff{1}Department of Mechanical and Aerospace Engineering, Monash University,\\Victoria 3800, Australia
\aff{2}Applied Mathematics Research Centre, Coventry University, Coventry, CV15FB, U.K}

\begin{document}

\maketitle

\begin{abstract}
This study seeks to elucidate the linear transient growth mechanisms in a uniform duct with square cross-section applicable to flows of electrically conducting fluids under the influence of an external magnetic field. A particular focus is given to the question of whether at high magnetic fields purely two-dimensional mechanisms exist, and whether these can be described by a computationally inexpensive quasi-two-dimensional model. Two Reynolds numbers of $5000$ and $15\,000$ and an extensive range of Hartmann numbers $0 \leq \Ha \leq 800$ were investigated. Three broad regimes are identified in which optimal mode topology and non-modal growth mechanisms are distinct. These regimes corresponding to low, moderate and high magnetic field strengths are found to be governed by the independent parameters; Hartmann number, Reynolds number based on the Hartmann layer thickness $R_H$, and Reynolds number built upon the Shercliff layer thickness $R_S$, respectively. Transition between regimes respectively occurs at $\Ha \approx 2$ and no lower than $R_H \approx 33.\dot{3}$. Notably for the high Hartmann number regime, quasi-two-dimensional magnetohydrodynamic models are shown to be an excellent predictor of not only transient growth magnitudes, but also the fundamental growth mechanisms of linear disturbances. This paves the way for a precise analysis of transition to quasi-two-dimensional turbulence at much higher Hartmann numbers than is currently achievable.
\end{abstract}

\section{Introduction}
This work focuses on the mechanisms driving transition to turbulence in duct flows subjected to strong magnetic fields. Because magnetic fields promote two-dimensionality at the expense of three-dimensional phenomena, transition to turbulence may follow an entirely different route than those found in hydrodynamics flows \citep{Sommeria:82}. In particular, a crucial question is whether a route to quasi-two-dimensional (Q2D) turbulence involving exclusively two-dimensional mechanisms exists. Besides a fundamental interest in finding new routes to turbulence, this question is also relevant to the design of nuclear fusion cooling blankets; namely, their heat and mass transport capabilities in environments subject to extreme magnetic fields \cite[$\sim 10$~T,][]{Smolentsev:07,Cassells:16}. 

Despite the intense Joule dissipation incurred in such conditions, duct elements can exhibit severe vibrations. The cause of which is likely due to the presence of Q2D turbulence producing large pressure fluctuations \citep{Alemany:79,Sommeria:82,Smolentsev:10}. The theoretical and experimental work by \citet{Sommeria:82}, \citet{Potherat:10} and \citet{Potherat:14} indicates that turbulence becomes two-dimensional when the ratio of Lorentz to inertial forces, measured by the so-called \textit{true} interaction parameter $N_t \equiv(\sigma B^2l_f/\rho U)(l_f/h)^2$, becomes sufficiently large (i.e. $N_t \gg 1$). Here $\sigma$ and $\rho$ are respectively the fluid's electrical conductivity and density, with $B$ and $U$ being the magnetic field and velocity magnitudes at a local length scale $l_f$, while $h$ is a characteristic field-aligned domain length scale (conventionally taken to be the duct width). 

Eliciting transition to Q2D turbulence can have a significant positive effect on heat transfer coefficients, and in turn, meeting stringent viability constraints of nuclear fusion reactor designs \citep{Cassells:16}. The question is then whether the path to such a state necessarily involves a breakdown phase of three-dimensional turbulence, or if a direct route from the Q2D laminar state to Q2D turbulence is possible. Mechanism of this type may also play a role in other flows with a tendency to two-dimensionality. Including metallurgical applications, flows with background rotation and where stratification is present (e.g. geophysical flows) \citep{Greenspan:68,Paret:97,Muller:Text}. The question of dimensionality of transition mechanisms in these frameworks is key to understanding the dynamics of all such systems; for example, the formation of atmospheric patterns.

Investigating transition to turbulence starts with the search for perturbations amplified by the flow dynamics. In shear flow profiles without inflexion points, the transition is generally subcritical and triggered by finite amplitude perturbations, some of which may arise from large transient amplification of initially infinitesimally small perturbations \citep{Schmid:01, Naraigh:15}. Thus, this paper is concerned with the linear transient growth of wall-bounded magnetohydrodynamic (MHD) flow subject to an externally applied magnetic field. The associated mechanisms are likely candidates for causing flow destabilisation, and subsequently bypass transition in both hydrodynamic and MHD flows \citep{Boberg:88,Trefethen:93,Reddy:93, Waleffe:97, Reshotko:01,  Biau:04, Krasnov:04}.

To date, the transient growth of perturbations in MHD channel or duct flows has been tackled either with a fully three-dimensional or a Q2D approach. Full three-dimensional analyses were conducted by \citet{Varet:02}, \citet{Airiau:04} and \citet{Krasnov:04} on the simpler problem of Hartmann flow (i.e. a channel flow between two perpendicular walls subjected to a magnetic field in the spanwise direction). However, the higher computational cost of resolving the thin Hartmann boundary layers confined these studies to relatively low magnetic fields. The thickness of these boundary layers scales as $\Ha^{-1}$, where $\Ha \equiv B h(\sigma/\rho\nu)^{1/2}$ is the Hartmann number (whose square represents the ratio of Lorentz and viscous forces), with $\nu$ being the fluid's kinematic viscosity. Nevertheless, subsequent direct numerical simulations (DNS) of Hartmann flows where a perturbation that maximise the linear amplification were added, recovered the critical regime parameters for the transition to turbulence found in experiments \citep{Moresco:04,Krasnov:04,Zienicke:2005}. The more realistic problem of MHD duct flow was subsequently tackled by \citet{Krasnov:10} employing a similar approach, with the same limitations imposed by computational cost, albeit more severe due to the presence of the Shercliff boundary layers (which form along walls parallel to the field) of thickness $\sim \Ha^{-1/2}$. Nevertheless, despite the important role of the Shercliff layers at high Hartmann number, none of the regimes investigated showed Q2D turbulence, and even less of a mechanism leading up to it. This is most likely because a sufficiently high Hartmann number could not be reached, where previous experiments have suggested values of $O(10^{3}$ - $10^{4})$ are needed \citep{Potherat:14, Baker:18}. 

The second Q2D approach, by contrast, specifically targets high-$\Ha$ regimes. It is based on the phenomenology that all scales are larger than the scale at which momentum diffusion along the magnetic field by the Lorentz force is balanced by inertia $l_c\simeq N_t^{1/3}$. Hence, if no length scale $l<l_c$ exists in either the laminar base flow or the turbulent state, then the flow can be described by two-dimensional dynamics. The governing equations are then obtained by averaging the full three-dimensional equations. The resulting shallow water model first derived by \citet{Sommeria:82}, hereafter SM82, consists of the two-dimensional incompressible Navier--Stokes equations with added linear friction accounting for the dissipative effects of the Hartmann layer. In this framework, instabilities in the Shercliff layers drive significant subcritical transient growth, albeit with energy gains often an order of magnitude less than in three-dimensional flow \citep{Potherat:07}. Despite evidence that SM82 can reproduce complex flows such as cylinder wakes when Q2D-scales exist, the question remains whether fine properties such as transient growth mechanisms are also accurately modelled \citep{Dousset:08, Ahmad:15}.

To address this issue, the present work aims to conduct a transient growth analysis of the MHD duct flow problem at high enough Hartmann numbers to reach the possible Q2D regimes accurately described by SM82. The key in this approach is to reach sufficiently high values of $\Ha$ whilst keeping the computational cost amenable to a full three-dimensional approach with adequate numerical precision. This delicate compromise is precisely the remit of spectral element methods, which shall form the basis of the numerical methods. The specific questions addressed are;

\begin{enumerate}[leftmargin=*,labelindent=16pt,itemindent=0pt,labelsep=0.2cm,label={(\arabic*)}]
\item Do purely Q2D transient growth mechanisms exist at sufficiently high but physically realistic Hartmann numbers?\vskip 0.3em
\item If yes, can these be accurately captured by means of SM82?\vskip 0.3em
\item What are the transitional regimes between the three-dimensional growth mechanisms discovered by \citet{Krasnov:10} and those found in the Q2D regime?
\end{enumerate}

The stake of establishing the validity of SM82 at sufficiently high Hartmann number are firstly, that the
transition to turbulence in such extreme regimes as those relevant to nuclear fusion could then be 
entirely understood by means of a computationally affordable Q2D approach; and secondly, a fully Q2D 
scenario for transition to turbulence would be at hand. 

This paper is structured as follows.  The problem is formulated in \S~\ref{prob_form} with the numerical methodology subsequently given in \S~\ref{num_meth}. In \S~\ref{trans_grow} the transient growth analysis and MHD model comparison are outlined and discussed. Concluding remarks are presented in \S~\ref{conc}.

\section{Problem Formulation}
\label{prob_form}
\begin{figure}
\centering
\begin{tabular}{l}
\begin{subfigure}{0.5\textwidth}
\begin{tikzpicture}
	%%% Edit the following coordinate to change the shape of your
	%%% cuboid

	%% Vanishing points for perspective handling
	\coordinate (P1) at (-7cm,1.5cm); % left vanishing point (To pick)
	\coordinate (P2) at (8cm,1.5cm); % right vanishing point (To pick)

	%% (A1) and (A2) defines the 2 central points of the cuboid
	\coordinate (A1) at (0em,0cm); % central top point (To pick)
	\coordinate (A2) at (0em,-2cm); % central bottom point (To pick)

	%% (A3) to (A8) are computed given a unique parameter (or 2) .8
	% You can vary .8 from 0 to 1 to change perspective on left side
	\coordinate (A3) at ($(P1)!.8!(A2)$); % To pick for perspective
	\coordinate (A4) at ($(P1)!.8!(A1)$);
	
	\coordinate (n1) at (-7.1cm,1.35cm); % left vanishing point (To pick)
	\coordinate (a2) at (-0.1cm,-2.15cm);
	\coordinate (a3) at ($(n1)!.8!(a2)$);
	
	\coordinate (n2) at (0.15cm,10cm); % left vanishing point (To pick)
	\coordinate (a9) at (-1.58cm,-1.28cm);
	\coordinate (a10) at (-1.58cm, 0.28cm);

	% You can vary .8 from 0 to 1 to change perspective on right side
	\coordinate (A7) at ($(P2)!.7!(A2)$);
	\coordinate (A8) at ($(P2)!.7!(A1)$);

	%% Automatically compute the last 2 points with intersections
	\coordinate (A5) at
	  (intersection cs: first line={(A8) -- (P1)},
			    second line={(A4) -- (P2)});
	\coordinate (A6) at
	  (intersection cs: first line={(A7) -- (P1)},
			    second line={(A3) -- (P2)});

	%%% Depending of what you want to display, you can comment/edit
	%%% the following lines
	%%% Drawing Coordinate System

	%% Possibly draw back faces

	\fill[gray!55,opacity=1] (A2) -- (A3) -- (A6) -- (A7) -- cycle; % face 6
	\node at (barycentric cs:A2=1,A3=1,A6=1,A7=1) {\tiny};
	
	\fill[gray!55,opacity=1] (A3) -- (A4) -- (A5) -- (A6) -- cycle; % face 3
	\node at (barycentric cs:A3=1,A4=1,A5=1,A6=1) {\tiny};
	
	\fill[gray!55,opacity=1] (A5) -- (A6) -- (A7) -- (A8) -- cycle; % face 4
	\node at (barycentric cs:A5=1,A6=1,A7=1,A8=1) {\tiny };
	
%	\draw[red,thick] (A5) -- (A6);
	\draw[black,thick] (A3) -- (A6);
%	\draw[thick,dashed] (A7) -- (A6);

	%% Possibly draw front faces

	\fill[gray!20,opacity=1] (A1) -- (A8) -- (A7) -- (A2) -- cycle; % face 1
	% \node at (barycentric cs:A1=1,A8=1,A7=1,A2=1) {\tiny f1};
	% \fill[red!50,opacity=0] (A1) -- (A2) -- (A3) -- (A4) -- cycle; % f2
	\node at (barycentric cs:A1=1,A2=1,A3=1,A4=1) {\tiny};
	\fill[gray!35,opacity=1] (A1) -- (A4) -- (A5) -- (A8) -- cycle; % f5
	\node at (barycentric cs:A1=1,A4=1,A5=1,A8=1) {\tiny};

	%% Possibly draw front lines
	\draw[thick] (A1) -- (A2);
	\draw[thick] (A3) -- (A4);
	\draw[thick] (A7) -- (A8);
	\draw[thick] (A1) -- (A4);
	\draw[thick] (A1) -- (A8);
	\draw[thick] (A2) -- (A3);
	\draw[doublearr] (a2) -- (a3) node[midway,below,yshift=0cm,xshift=-0.2cm] {$2a$};
    \draw[doublearr] (a9) -- (a10) node[midway,below,yshift=0.25cm,xshift=-0.3cm] {$2a$};
	\draw[thick] (A2) -- (A7);
	\draw[thick] (A4) -- (A5);
	\draw[thick] (A8) -- (A5);
	
	% Drawing B vector
	\draw[black,opacity=1,line width=2.2pt, xshift=0.65cm, yshift=0.5cm,arrows={-Stealth[scale=1.0]}] (0,-3.2,-3) -- (0,-1.5,-3) node[black,at start, left, xshift=0.7cm, yshift=1.0cm]{$\boldsymbol{B_0}$};

\node[xshift=1.85cm,yshift=0.3cm,label=above:$$]  (vh) at (-1.05,-1.18)  {$\BU_0$};
%\node[xshift=1.8cm,yshift=1.5cm,label=above:$$]  (vh) at (-1.05,-1.18)  {$\BU_0$};

\draw[thick,xshift=1.23cm, yshift=-1.85cm,arrows={-Latex[length=1.1mm,width=1.1mm]}] (-2,1,0) -- (-1.4,1.15,0) node[anchor=south]{$z$};
	\draw[line width=0.7pt,xshift=1.23cm, yshift=-1.85cm,arrows={-Latex[length=1.1mm,width=1.1mm]}] (-2,1,0) -- (-2,1.6,0) node[anchor=east]{$y$};
	\draw[thick,xshift=1.23cm, yshift=-1.85cm,arrows={-Latex[length=1.1mm,width=1.1mm]}] (-2,1,0) -- (-1.2,1,0.65) node[anchor=north,xshift=-0.1cm]{$x$};

\node[xshift=1.23cm, yshift=-1.85cm,label=above:$$,scale=0.75]  (x0) at (-2,1,0)  {$\bullet$};

% Hartann Nodes
%  \node at (0,0) {$F : I \times I \rightarrow X$};
  \node[xshift=1.4cm,yshift=-0.28cm,label=below:$$]  (x1) at (-1.4,0.28)  {$$};
  \node[xshift=1.4cm,yshift=-0.7cm,label=above:$$]  (x0) at (-1.4,-1.29)  {$$};
  \node[xshift=1.4cm,yshift=-0.7cm,label=above:$$]  (x2) at (-0.86,-0.515)  {$$};
  \node[xshift=1.32cm,yshift=-0.21cm,label=above:$$]  (x3) at (-1.1,0.23)  {$$};

  \node[xshift=1.4cm,yshift=-0.64cm,label=above:$$]  (x4) at (-1.05,-1.18)  {$$};
  \node[xshift=1.39cm,yshift=-0.18cm,label=above:$$]  (x5) at (-0.86,0.1)  {};
  \node[xshift=1.4cm,yshift=-0.61cm,label=above:$$]  (x6) at (-0.86,-1)  {};

  %\node  at (0,0)  {$\subset X$};
  \draw (x1.center) to [out=0,in=180](x3.center);
  \draw (x3.center) to [out=5,in=90](x5.center);

  \draw (x0.center) to [out=20,in=205](x4.center);
  \draw (x4.center) to [out=30,in=-90](x6.center);

  \draw (x5.center) to [out=-88,in=90](x2.center);
  \draw (x6.center) to [out=88,in=-90](x2.center);

 % Hartmann Arrows
 \node[xshift=1.4cm,yshift=-0.28cm,label=below:$$]  (eh1) at (-1.4,0.28)  {$$};

 \node[xshift=1.4cm,yshift=-0.52875cm,label=below:$$]  (eh2) at (-1.4,0.28)  {$$};

  \node[xshift=1.4cm,yshift=-0.7775cm,label=below:$$]  (eh3) at (-1.4,0.28)  {$$};

   \node[xshift=1.4cm,yshift=-1.02625cm,label=below:$$]  (eh4) at (-1.4,0.28)  {$$};

  \node[xshift=1.4cm,yshift=-1.275cm,label=below:$$]  (eh5) at (-1.4,0.28)  {$$};

  \node[xshift=1.4cm,yshift=-1.52375cm,label=below:$$]  (eh6) at (-1.4,0.28)  {$$};

  \node[xshift=1.4cm,yshift=-1.7725cm,label=below:$$]  (eh7) at (-1.4,0.28)  {$$};
    \node[xshift=1.4cm,yshift=-2.02125cm,label=below:$$]
     (eh8) at (-1.4,0.28)  {$$};

\node[xshift=1.4cm,yshift=-0.7cm,label=above:$$]  (eh9) at (-1.4,-1.29) {$$};

\node[xshift=1.94cm,yshift=-0.42975cm,label=below:$$]  (bh2) at (-1.4,0.28)  {$$};

  \node[xshift=1.94cm,yshift=-0.6695cm,label=below:$$]  (bh3) at (-1.4,0.28)  {$$};

   \node[xshift=1.94cm,yshift=-0.91925cm,label=below:$$]  (bh4) at (-1.4,0.28)  {$$};

  \node[xshift=1.94cm,yshift=-1.149cm,label=below:$$]  (bh5) at (-1.4,0.28)  {$$};

  \node[xshift=1.94cm,yshift=-1.38375cm,label=below:$$]  (bh6) at (-1.4,0.28)  {$$};

  \node[xshift=1.94cm,yshift=-1.6125cm,label=below:$$]  (bh7) at (-1.4,0.28)  {$$};

  \node[xshift=1.94cm,yshift=-1.83525cm,label=below:$$]
     (bh8) at (-1.4,0.28)  {$$};

\draw[thick,arrows={-Latex[length=1.1mm,width=1.1mm]}] (eh2.center)-- (bh2.center);

\draw[thick,arrows={-Latex[length=1.1mm,width=1.1mm]}] (eh3.center)-- (bh3.center);

\draw[thick,arrows={-Latex[length=1.1mm,width=1.1mm]}] (eh4.center)-- (bh4.center);

\draw[thick,arrows={-Latex[length=1.1mm,width=1.1mm]}] (eh5.center)-- (bh5.center);

\draw[thick,arrows={-Latex[length=1.1mm,width=1.1mm]}] (eh6.center)-- (bh6.center);

\draw[thick,arrows={-Latex[length=1.1mm,width=1.1mm]}] (eh7.center)-- (bh7.center);

\draw[thick,arrows={-Latex[length=1.1mm,width=1.1mm]}] (eh8.center)-- (bh8.center);

  % Shercliff Profile
  \node[xshift=1.4cm,yshift=-0.28cm,label=below:$$]  (x10) at (-1.4,0.28)  {$$};
  \node[xshift=1.6cm,yshift=-0.05cm,label=above:$$]  (x11) at (-1.4,0.28)   {$$};
    \node[xshift=1.4cm,yshift=0.03cm,label=above:$$]  (x12) at (-1.4,0.28)  {$$};

  \draw (x10.center) to [out=15,in=-40](x11.center);
  \draw (x11.center) to [out=150,in=-15](x12.center);

  \node[xshift=0cm,yshift=0.02cm,label=below:$$]  (x13) at (-1.4,0.28)  {$$};  \node[xshift=1.15cm,yshift=0.08cm,label=above:$$]  (x14) at (-1.4,0.28)   {$$};
 \draw (x13.center) to [out=5,in=175](x14.center);
  \draw (x14.center) to [out=-5,in=165](x12.center);

% Shercliff Arrows

\node[xshift=1.28cm,yshift=-0.26cm,label=below:$$]  (es2) at (-1.4,0.28)  {$$};

\node[xshift=1.065cm,yshift=-0.21cm,label=below:$$]  (es3) at (-1.4,0.28)  {$$};

\node[xshift=0.77cm,yshift=-0.14cm,label=below:$$]  (es4) at (-1.4,0.28)  {$$};

\node[xshift=0.5cm,yshift=-0.08cm,label=below:$$]  (es5) at (-1.4,0.28)  {$$};

\node[xshift=0.2cm,yshift=-0.02cm,label=below:$$]  (es6) at (-1.4,0.28)  {$$};

\node[xshift=1.58cm,yshift=-0.19cm,label=below:$$]  (bs2) at (-1.4,0.28)  {$$};

\node[xshift=1.59cm,yshift=-0.06cm,label=below:$$]  (bs3) at (-1.4,0.28)  {$$};

\node[xshift=1.43cm,yshift=0.02cm,label=below:$$]  (bs4) at (-1.4,0.28)  {$$};

\node[xshift=1.15cm,yshift=0.065cm,label=below:$$]  (bs5) at (-1.4,0.28)  {$$};

\node[xshift=0.8cm,yshift=0.075cm,label=below:$$]  (bs6) at (-1.4,0.28)  {$$};

\draw[thick,arrows={-Latex[length=1.1mm,width=1.1mm,scale=1]}] (es2.center)-- (bs2.center);

\draw[thick,arrows={-Latex[length=1.1mm,width=1.1mm]}] (es3.center)-- (bs3.center);

\draw[thick,arrows={-Latex[length=1.1mm,width=1.1mm]}] (es4.center)-- (bs4.center);

\draw[thick,arrows={-Latex[length=1.1mm,width=1.1mm]}] (es5.center)-- (bs5.center);

\draw[thick,arrows={-Latex[length=1.1mm,width=1.1mm]}] (es6.center)-- (bs6.center);
\end{tikzpicture}
\end{subfigure}

\begin{subfigure}{.3\textwidth}
\begin{tikzpicture}
\node[anchor=south west,inner sep=0] at (0,0){\includegraphics[width=\textwidth]{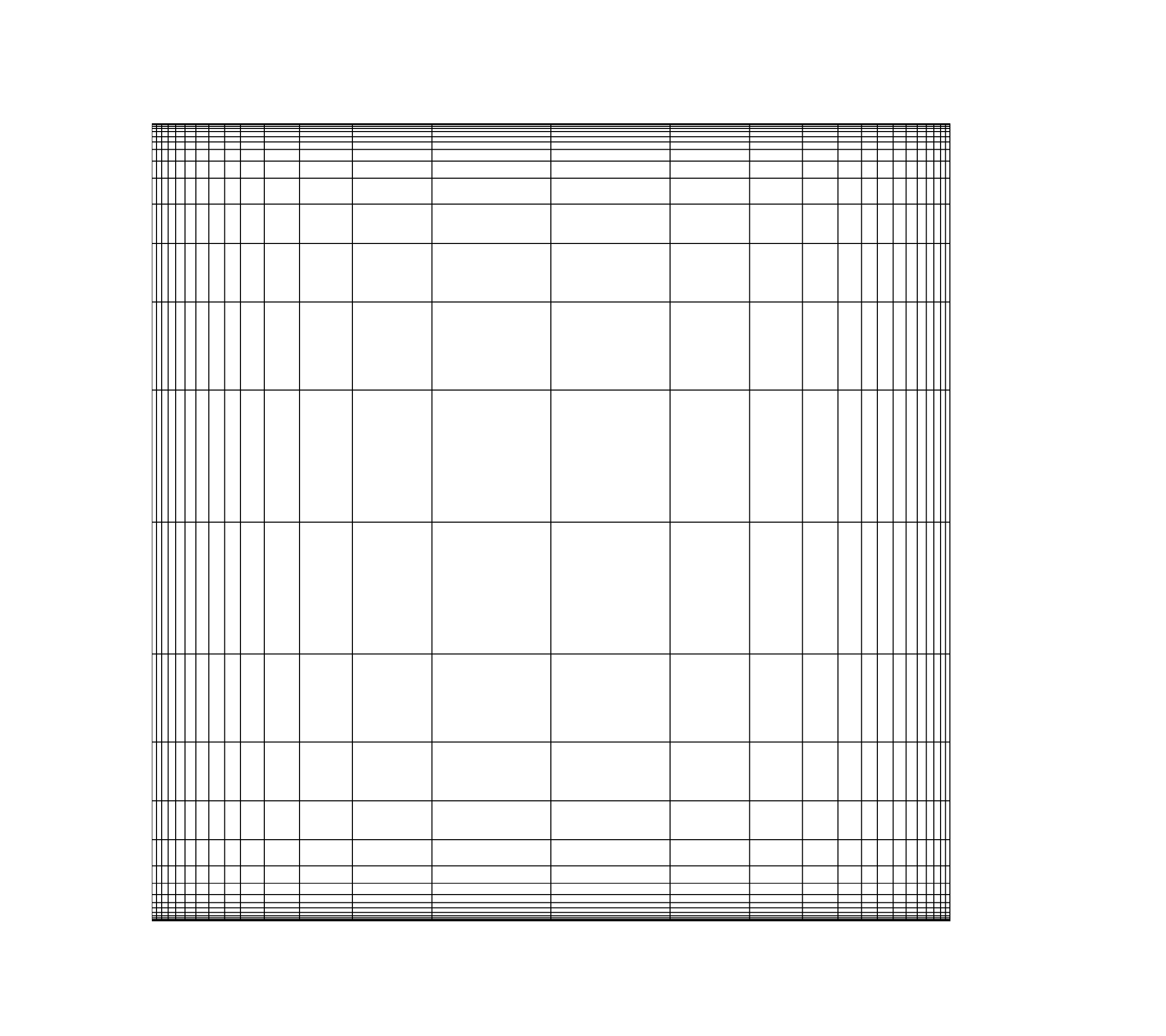}};
\draw[line width=0.5pt,xshift=3.92cm, yshift=0.79cm,arrows={-Latex[length=1.1mm,width=1.1mm]}] (-2,1,0) -- (-2,1.6,0) node[anchor=east]{$y$};
	\draw[line width=0.5pt,xshift=3.92cm, yshift=0.79cm,arrows={-Latex[length=1.1mm,width=1.1mm]}] (-2,1,0) -- (-1.4,1,0) node[anchor=north,xshift=-0.1cm]{$x$};
	\node[xshift=3.92cm, yshift=0.79cm,label=above:$$,scale=0.4]  (x0) at (-2,1,0)  {$\bullet$};
\end{tikzpicture}
\end{subfigure}

\end{tabular}
\caption{Flow configuration for the straight duct with square cross-section (left) and an example graded spectral element distribution in the $x$-$y$ plane for three-dimensional investigations at $\Ha=30$. The Shercliff and Hartmann boundary layers respectively develop on the vertical and horizontal sidewalls.}
\label{schematic}
\end{figure}
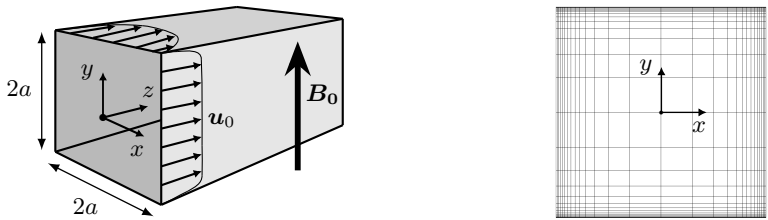

A fluid with electrical conductivity $\sigma$, kinematic viscosity $\nu$ and density $\rho$, flows through a square duct with cross-sectional width $2a$ as depicted in figure~\ref{schematic}. The electrically insulated vertical and horizontal duct walls are respectively located at $x=\pm a$ and $y=\pm a$. 
An external homogeneous magnetic field $\boldsymbol{B_0} = B_0 \boldsymbol{e_y}$ is imposed in the vertical direction. A fully developed velocity profile is prescribed at the inlet, and a constant streamwise pressure gradient is applied to drive the flow through the duct.

The MHD governing equations are written in the quasi-static, low-magnetic Reynolds number ($R_m$) approximation. 
In this approximation, the motion of the conducting fluid in the magnetic field induces electric currents that are non-negligible, however, the magnetic-field induced by these currents is. Therefore, the agglomerated  magnetic field remains indistinguishable from $\bm{B_0}$, and the magnetic field transport equation is not needed. Non-dimensionalisation of the governing equations is achieved by taking the scale transformations for length $a$, velocity $U_0$ (where $U_0$ is the peak inlet velocity), pressure $\rho U_0^2$, time $a/U_0$, magnetic field strength $B_0$ and lastly, for the electric potential $a U_0 B_0$. It follows that the dimensionless quasi-static momentum and continuity equations can be written as
\begin{subequations}
\label{NumIntro:NSESum}%
\begin{gather}
\frac{\partial \BU}{\partial t}  = -(\BU \bcdot \BN )\BU - \BN  p +  \frac{1}{Re} \nabla^2 \BU + \frac{{Ha}^2}{\Rey}(\boldsymbol{j}\times \EY),\\[3pt]
\BN  \bcdot \BU = 0,
\end{gather}
\end{subequations}
where the electric current density vector $\boldsymbol{j}$ is given by Ohm's law
\BEQ
\label{Ohm}
\boldsymbol{j}=-\BN  \phi + \BU \times \EY.
\EEQ
Here $\BU(x,y,z,t)$ is the time $t$ dependent velocity vector field having Cartesian $x$-, $y$- and $z$-components $u$, $v$ and $w$, while $\phi(x,y,z,t)$ and $p(x,y,z,t)$ are the scalar electric potential and pressure fields, respectively. The dimensionless groups $Re\equiv U_0a/\nu$ and $\Ha \equiv aB_0\sqrt{\sigma/\rho\nu}$ are respectively the Reynolds number and the Hartmann number. 

In the present work, Hartmann numbers between $ 0 \leq \Ha \leq 800$ are investigated, which significantly extends the range covered by \citet{Krasnov:10} and helps bridge the gap between three-dimensional and Q2D models for transient growth of linear perturbations. Primarily two Reynolds numbers are investigated; $\Rey =5000$ and $\Rey =15\,000$. The former facilitates comparisons with existing literature, and is below the exponential instability limit found for hydrodynamic Poiseuille flow; and the latter elucidates the effect of $\Ha$ and $\Rey$ on transient growth and related mechanisms. In addition to these two Reynolds numbers, supplementary cases at $\Rey=2000$ and $10\,000$ are also conducted to support findings in \S~\ref{trans_grow}.

Utilising the low-$R_m$ approximation and by assuming an electrically neutral fluid, the current density can be considered solenoidal, i.e.\ $\BN  \bcdot \boldsymbol{j} = 0$. By taking the divergence of \eqref{Ohm}, a Poisson equation for the electric potential is formed,
\BEQ
\label{elec_pot}
\nabla^2\phi = \BN  \bcdot \left(\BU \times \EY \right).
\EEQ
No-slip conditions are defined at the walls through Dirichlet-type boundary conditions
\BEQ
\label{noslip}
\BU = 0 \text{~~at~} x,y = \pm a,
\EEQ
with electrically insulated horizontal and vertical side-walls respectively imposed by the Neumann-type boundary conditions
\begin{subeqnarray}
\label{elec_ins}
\frac{\partial \phi}{\partial x} & = & 0 \text{~~at~} x = \pm a,\\[3pt]
\frac{\partial \phi}{\partial y} & = & 0 \text{~~at~} y = \pm a.
\end{subeqnarray}
\subsection{Quasi-two-dimensional model}
The core flow dynamics can be suitably approximated through the use of the quasi-two-dimensional MHD model proposed by \citet{Sommeria:82} in the limit of high interaction parameter ($N \equiv\Ha^2/\Rey \gg 1$) and Hartman number ($\Ha \gg 1$). Electrically insulated side-walls are assumed in the construction of the model, which leads to a two-dimensional Navier--Stokes equation augmented via a linear braking term representing friction due to the Hartmann layers. By employing the subscript $\perp$ to represent the horizontal components of the averaging of velocity and pressure in the $y$-direction, in addition to the horizontal trace of operators, the quasi-two-dimensional MHD equations are given as
\begin{subequations}
\label{SM82}%
\begin{gather}
\frac{\partial \BU_{\perp}}{\partial t}  = - \left(\BU_{\perp} \bcdot \BN_\perp \right) \BU_{\perp} - \BN_\perp  p_\perp +  \frac{1}{Re} \nabla_\perp^2 \BU_{\perp} - \frac{H}{\Rey}\BU_{\perp},\\[3pt]
\bnabla\bcdot\BU_{\perp} = 0,
\end{gather}
\end{subequations}
where $\BU_{\perp}(x,z,t)$ is a two-dimensional time-dependent velocity vector field and $H \equiv n\Ha(a/b)^2$ is a Hartmann friction parameter representing the effect of the Lorentz force on the flow, where $n$ represents the number of Hartmann layers (e.g.\ $n=2$ for the present configuration) and $b$ is the out-of-plane duct depth. Consistent with the three-dimensional model, no-slip boundary conditions are imposed on the vertical walls such that
\BEQ
\label{elec_ins_q2d}
\BU_{\perp} = 0 \text{~~at~} x = \pm a.
\EEQ

\subsection{Base flow and linear perturbations}
\label{sec:BF_LP}
\subsubsection{Three-dimensional flows}
The three-dimensional MHD analysis employs a base flow that is steady, streamwise-independent $\BU_0(x,y) = w_0(x,y)$ and which conforms to the boundary conditions \eqref{noslip}--\eqref{elec_ins}. The solution is obtained through time integration of the MHD governing equations \eqref{NumIntro:NSESum} to a time-independent steady-state flow. The general solutions for the velocity, electric-potential and pressure fields can be represented by the summation of the base flow (denoted by the subscript $0$) and perturbation equations (denoted by the superscript $\prime$) such that
\BEQ
\label{gen_base_pert}
\left[\BU,\phi,p\right] =\left[\BU_0,\phi_{0},p_0\right] + \left[\BU^\prime,\phi^\prime,p^\prime\right].
\EEQ

The linearised governing MHD equations are constructed by substituing \eqref{gen_base_pert} into \eqref{NumIntro:NSESum} and \eqref{elec_pot}, whilst retaining only the lowest-order linear terms. The resulting equations are
\begin{subequations}
\label{NumIntro:linear_MHD}
\begin{gather}
\frac{\partial \BU^\prime}{\partial t} = -\mathrm{DN}^\prime\BU^\prime - \BN  p^\prime +  \frac{1}{\Rey} \nabla^2 \BU^\prime + \frac{{Ha}^2}{\Rey}\left[\left(-\BN  \phi^\prime + \BU^\prime \times \EY\right)\times\EY\right],\label{3Da}\\[3pt]
\bnabla\bcdot\BU^\prime = 0,\\[3pt]
\nabla^2\phi^\prime = \BN  \bcdot \left(\BU^\prime \times \EY \right),
\end{gather}
\end{subequations}
where $\mathrm{DN}^\prime\BU^\prime = \left(\BN \BU_{0}\right)\bcdot \BU^\prime + (\BU_{0} \bcdot \BN )\BU^\prime$ is the linearised advection operator.

The proceeding transient growth analysis requires the construction of the adjoint form of the linearised governing MHD equations. Through an analogous derivation to that presented in \citet{Blackburn:08}, the adjoint reformulation of \eqref{NumIntro:linear_MHD} is given as
\begin{subequations}
\label{NumIntro:linear_MHD_adj}
\begin{gather}
-\frac{\partial \BU^\ast}{\partial t} = -\mathrm{DN}^\ast\BU^\ast - \BN  p^\ast +  \frac{1}{\Rey} \nabla^2 \BU^\ast + \frac{{Ha}^2}{\Rey}\left[\left(-\BN  \phi^\ast + \BU^\ast \times \EY\right)\times\EY\right],\label{3DAa}\\[3pt]
\bnabla\bcdot\BU^\ast = 0,\\[3pt]
\nabla^2\phi^\ast = \BN  \bcdot \left(\BU^\ast \times \EY \right),
\end{gather}
\end{subequations}
where $\mathrm{DN}^\ast\BU^\ast = \left(\BN \BU_{0}\right)^\text{T}\bcdot \BU^\ast - (\BU_{0} \bcdot \BN )\BU^\ast$ is the adjoint linear advection operator. $\BU^\ast$, $p^\ast$ and $\phi^\ast$ denote the velocity, pressure and electro-potential fields of the adjoint perturbation scenario, respectively.

Spatial homogeneity in the streamwise direction is assumed so that the time dependent three-dimensional linear perturbations for both the forward and adjoint systems are considered in the form of the respective decoupled orthogonal Fourier modes
\BEQ
\left[\BU^\prime,\phi^\prime,p^\prime\right]= \left[\hat{u},\hat{v},\hat{w},\hat{\phi},\hat{p}\right](x,y,t)\bcdot e^{ik z},
\EEQ
\BEQ
\left[\BU^\ast,\phi^\ast,p^\ast\right]= \left[\hat{u}^\ast,\hat{v}^\ast,\hat{w}^\ast,\hat{\phi}^\ast,\hat{p}^\ast\right](x,y,t)\bcdot e^{ik z},
\EEQ
where $k = 2\pi/\ell_z$ is the associated streamwise wavenumber in the $z$-direction (a parameter to be varied in this study), and $\ell_z$ is the periodicity length of the domain, also in the $z$-direction. The circumflex ($\hat{\bcdot}$) represents the Fourier coefficient of the associated base variable.
\subsubsection{Quasi-two-dimensional flows}
\label{sec:q2d_num_meth}
Unlike the three-dimensional transient growth analysis, the method employing the SM82 model does not utilise Fourier modal expansion in the third dimension, but instead discretises the domain using spectral elements. For this case, the analytical solution for a one-dimensional steady Q2D baseflow
\BEQ
\label{aa_q2d}
{\BU}_{\perp,0}(x) = \frac{\cosh\sqrt{H}}{\cosh\sqrt{H}-1}\left(1-\frac{\cosh(\sqrt{H}x)}{\cosh\sqrt{H}}\right),
\EEQ
as defined in \citet{Potherat:07} is employed. Subjecting \eqref{aa_q2d} to two-dimensional time-dependent linear perturbations of the form
\BEQ
\label{q2d_perts}
\left[\BU_{\perp}^\prime,p_{\perp}^\prime\right]=\left[u^\prime,w^\prime,p^\prime\right]\left(x,z,t\right),
\EEQ
the flow can then be described by the general solution
\BEQ
\label{gen_q2d}
\left[\BU_{\perp},p_{\perp}\right]= \left[{\BU}_{\perp,0},{p}_{\perp,0}\right] + \left[\BU_{\perp}^\prime,p_{\perp}^\prime\right].
\EEQ
Hence, through substitution of \eqref{gen_q2d} into \eqref{SM82}, while keeping only the lowest-order linear terms, the linearised Q2D SM82 equations are given as
\begin{subequations}
\label{SM82_lin}
\begin{gather}
\frac{\partial \BU^\prime_{\perp}}{\partial t} = -\mathrm{DN}\BU^\prime_{\perp} - \BN_{\perp}  p^\prime_{\perp} +  \frac{1}{\Rey} \nabla_{\perp}^2 \BU^\prime_{\perp} + \frac{{H}}{\Rey}\BU^\prime_{\perp},\label{2Da}\\[3pt]
\bnabla_{\perp}\bcdot\BU^\prime_{\perp} = 0,
\end{gather}
\end{subequations}
where $\mathrm{DN}^\prime\BU^\prime_{\perp} = \left(\BN_{\perp} {\BU}_{\perp,0}\right)\bcdot \BU_{\perp}^\prime + ({\BU}_{\perp,0} \bcdot \BN_{\perp})\BU^\prime_{\perp}$ is the Q2D linearised advection term. Similarly, the adjoint linearised SM82 equations are given by
\begin{subequations}
\label{SM82_lin_adj}
\begin{gather}
-\frac{\partial \BU^\ast_{\perp}}{\partial t} = -\mathrm{DN}^\ast\BU^\ast_{\perp} - \BN_{\perp}  p^\ast_{\perp} +  \frac{1}{\Rey} \nabla_{\perp}^2 \BU^\ast_{\perp} + \frac{{H}}{\Rey}\BU^\ast_{\perp},\label{2DAa}\\[3pt]
\bnabla_{\perp}\bcdot\BU^\ast_{\perp} = 0,
\end{gather}
\end{subequations}
where $\mathrm{DN}^\ast\BU^\ast_{\perp} = \left(\BN_{\perp} {\BU}_{\perp,0}\right)^\text{T}\bcdot \BU_{\perp}^\ast - ({\BU}_{\perp,0} \bcdot \BN_{\perp} )\BU^\ast_{\perp}$ is the adjoint Q2D linearised advection term.

The streamwise periodicity length of the Q2D domain $\ell_{z,\perp}$ is determined from the three-dimensional transient growth optimal wavenumber $k_{opt}$ (as defined in \S~\ref{sec:TG}) at an equivalent $\Ha$ such that $\ell_{z,\perp} = 2\pi/k_{opt}$.
\subsection{Transient growth analysis}
\label{sec:TG}
The present work is interested in the transient energy growth of linear perturbations over a finite time interval $\tau$ (a parameter to be varied in this study). Outlined in this section is the methodology relating to the three-dimensional MHD case only. The formulation pertaining to the SM82 model is analogous to that presented next, and is therefore omitted for brevity. 

The governing direct and adjoint \thd MHD linearised equations, given respectively in \eqref{NumIntro:linear_MHD} and \eqref{NumIntro:linear_MHD_adj}, are solved subject to the boundary conditions \eqref{noslip}--\eqref{elec_ins} in combination with suitable initial conditions. The nature of linear transient growth is determined through a direct-adjoint eigenvalue system which is solved using a methodology consistent with \citet{Barkley:08}. In this direct-adjoint system, an initial perturbation field providing a given gain in kinetic energy is found through the construction of an auxiliary eigenvalue problem
\BEQ
\label{eig_prob}
\mathscr{A}^\ast(\tau)\mathscr{A}(\tau)\hat{\BU}_k = \lambda_k\hat{\BU}_k, \qquad \Vert\hat{\BU}_k\Vert = 1,
\EEQ
where $\lambda_k$ and $\hat{\BU}_k$ denote eigenvalues and normalised eigenvectors, respectively. The state-transition operator $\mathscr{A}(\tau)$ represents the terms on the right hand side of \eqref{3Da}. Therefore, the operator describes the time evolution of an arbitrary initial perturbation $\BU^\prime(0)$ to $t=\tau$ such that $\BU^\prime(\tau)=\mathscr{A}(\tau)\BU^\prime(0)$. Similarly,  $\mathscr{A}^{*}(\tau)$  is the equivalent adjoint evolution operator of $\mathscr{A}(\tau)$ that evolves an equivalent adjoint variable ${\BU}^{*}(\tau)$, as solved via \eqref{3DAa}, backwards in time from $t=\tau$ to $t=0$. An eigenvector $\hat{\BU}_k$ that describes the initial perturbation field $\BU^\prime(0)$ that generates growth $\lambda_k$ over time $\tau$ can then be found by  taking a singular value decomposition of $\mathscr{A}^\ast(\tau)\mathscr{A}(\tau)$. The real and orthonormal right singular vectors resulting from such a decomposition represent the eigenvectors $\hat{\BU}_k$  describing the initial perturbation field $\BU^\prime(0)$ which generates growth $\lambda_k$ over time $\tau$. The singular value decomposition is determined using an implicitly restarted Arnoldi iterative method, the particulars of which can be found in \citet{ARPACK:98}. It is important to note that the operator $\mathscr{A}^\ast(\tau)\mathscr{A}(\tau)$ is not explicitly constructed: instead, the action of $\mathscr{A}^\ast(\tau)\mathscr{A}(\tau)$ on the field $\BU^\prime$ is determined through forward time integration of an initial condition over a time interval $\tau$ via \eqref{NumIntro:linear_MHD}, and then subsequently backwards from $\tau$ to the initial time under the adjoint system in \eqref{NumIntro:linear_MHD_adj}.

For the present transient growth analysis, it is the eigenvector producing the maximal growth for all $k$ and $\tau$ that is of particular interest. This eigenvector is commonly referred to as the optimal mode and this terminology is thus used hereafter. The optimal energy gain $G_{\text{max}}$ occurs at an optimal time interval $\TO$ having optimal streamwise wavenumber $k_{\rm{opt}}$.
\section{Numerical Methodology}
\label{num_meth}
\subsection{Spatial discretisation}
A high-order spectral element method following that described in \citet{Karniadakis:91} is employed for spatial discretisation of the governing equations. The domain is meshed using quadrilateral macro-elements with internally applied Lagrangian polynomial functions, the order of which, $N_p$, is varied to control spatial resolution in the spanwise $x$-$y$ cross-plane for three-dimensional investigations, and in the streamwise $x$-$z$ plane for the Q2D case. Linear grading of macro-elements towards all solid surfaces is employed to help resolve regions that experience the largest flow gradients. The spacing of macro elements is scaled with respect to the $\Ha$-dependent boundary layer thickness; ensuring that a minimum of 8 macro elements span their height. For the Q2D analysis, a minimum of 20 macro-elements span the streamwise and spanwise directions. Examples of the macro-element distribution in the $x$-$y$ cross-plane for three-dimensional investigations at $\Ha =30$ is shown in figure~\ref{schematic}.

\subsection{Temporal discretisation}
\label{temp_disc}
A third-order backwards differentiation scheme involving an operator-splitting method is employed for temporal advancement of both three-dimensional and quasi-two-dimensional governing equations. This is analogous to the method described in \citet{Karniadakis:91}. To the best of the authors' knowlegde, this temporal discretisation scheme has not been previously implemented for three-dimensional MHD flows. The implicit solution of the electric potential field at each time step based on the high-order projection of the velocity field, circumvents common issues pertaining to charge conservation encountered in finite-volume-based schemes \citep{Ni:07a}. As such, presented here is a full outline for the non-linear three-dimensional governing MHD equations in \eqref{NumIntro:NSESum}. The three-dimensional linearised equations \eqref{NumIntro:linear_MHD} and \eqref{NumIntro:linear_MHD_adj}, as well as the corresponding SM82 equations in \eqref{SM82_lin} and \eqref{SM82_lin_adj} are solved analogously.

The MHD governing equations \eqref{NumIntro:NSESum} are evaluated at future time step ($r+1$), with the time derivative term approximated via backwards differentiation. The semi-discrete form of equation~\eqref{NumIntro:NSESum} is

\begin{multline}
\label{time_der}%
\frac{\alpha_0\BU^{(r+1)}-\sum^J_{q=1}\alpha_q\BU^{(r-q+1)}}{\Delta t} =  \\
 \left[ -\left(\BU \bcdot \BN \right)\BU -\BN p + \frac{1}{\Rey}\nabla^2 \BU +  \frac{{Ha}^2}{\Rey}\left[\left(-\BN  \phi + \BU \times \EY\right)\times\EY\right]\right]^{(r+1)},
\end{multline}
where $J$ denotes the order of the scheme (e.g.\ $J=3$ in the present formulation) and $\alpha_q$ are the corresponding coefficients (i.e.\ $\alpha_0=11/6$, $\alpha_1=-3$, $\alpha_2=3/2$ and $\alpha_3=-1/3$).

An operator-splitting approach is used to split the resolution of  equation~\eqref{time_der} into three sub-steps. As a precursor, an explicit projection of the velocity field to the future ($r+1$) time is evaluated,
\BEQ
\widetilde{\BU} = \sum^{J-1}_{q=0}\gamma_q\BU^{(r-q)},
\EEQ
where for $J=3$ the $\gamma_q$ coefficients are $\gamma_0=3$, $\gamma_1=-3$, $\gamma_2=1$. Introducing intermediate velocity fields $\BU^\dagger$ and $\BU^\ddagger$, the semi-discrete projection for the advection, pressure and viscous diffusion terms are then respectively given by
\begin{subequations}
\begin{gather}
\frac{\BU^\dagger-\sum^J_{q=1}\alpha_q\BU^{(r-q+1)}}{\Delta t}  =  -\left(\widetilde{\BU}\bcdot\BN\right)\widetilde{\BU} + \frac{{Ha}^2}{\Rey}\left[\left(-\BN\phi^{(r+1)}+\widetilde{\BU}\times\EY\right)\times\EY\right],\label{seca} \\[3pt]
\frac{\BU^\ddagger-\BU^\dagger}{\Delta t}  =  -\BN p^{(r+1)},\label{secb} \\[3pt]
\frac{\alpha_0\BU^{(r+1)}-\BU^\ddagger}{\Delta t}  =  \frac{1}{\Rey}\nabla^2\BU^{(r+1)}.\label{secc}
\end{gather}
\end{subequations}
The electric potential field in equation~\eqref{seca} is determined from a Poisson equation constructed from equation~\eqref{elec_pot} while enforcing the divergence-free constraint on the electric current field,
\BEQ
\label{eq:phi-poisson}
\nabla^2 \phi^{(r+1)} = \BN\bcdot\left(\widetilde{\BU}\times\EY\right).
\EEQ

Taking the divergence of equation~\eqref{secb} and enforcing the divergence-free constraint on the second intermediate velocity field (i.e.\,$\bnabla \bcdot \BU^\ddagger = 0$) yields a Poisson equation for the pressure,
\BEQ
\label{eq:pressure-poisson}
\nabla^2 p^{(r+1)} = \bnabla \bcdot \left(\frac{\BU^\dagger}{\Delta t}\right).
\EEQ
Equation~\eqref{secc} is recast as a Helmholtz equation,
\BEQ
\label{eq:vel-helmholtz}
\left(\nabla^2-\alpha_0\frac{\Rey}{\Delta t}\right)\BU^{(r+1)} = -\frac{\Rey}{\Delta t}\BU^\ddagger,
\EEQ
for solution of the final velocity field $\BU^{(r+1)}$.

Dirichlet and Neumann boundary conditions on $\phi$, $p$ and $\BU$ are respectively imposed through equations~\eqref{eq:phi-poisson}--\eqref{eq:vel-helmholtz}. In addition to the boundary conditions defined in \eqref{noslip}--\eqref{elec_ins}, the above equations are solved subject to high order pressure field Neumann-type boundary conditions enforced on all domain perimeters. This allows for third-order time accuracy to be preserved \citep{Karniadakis:91}.

The present numerical scheme is built upon an existing solver that has been rigorously validated for two- and three-dimensional incompressible Navier--Stokes flows and Q2D MHD flows \citep{Hussam:12a, Hussam:12b, Hussam:13, Vo:15, Cassells:16, Hamid:16, Leigh:16, Ng:16, Sheard:16}. Nonetheless, a careful eye was kept on the conservation properties of the numerical algorithm; specifically, the discretised divergence of the velocity and current fields. The domain-integrated norms for both values did not exceed $10^{-9}$ in any of the simulations presented herein.

\subsection{Numerical validation}
To ensure an accurate computation of the three-dimensional and quasi-two-dimensional steady MHD baseflows, the velocity profiles were validated against analytical solutions as developed by \citet{Hunt:65} and \citet{Potherat:07}, respectively. The integrated error between the numerical and analytical velocity field solutions was found to be below 0.001\% of the peak velocity for all cases presented here. Furthermore, convergence of wall shear stresses $\tau_{zx}$ and $\tau_{zy}$, the standard $\mathcal{L}^2$ norm, and current magnitude $|j|$ of better than 0.03\% (where 0\% convergence is exact) was satisfied for polynomial orders of $N_p=8$; which is thus employed hereafter. It is worth noting here that adequate convergence was obtained for domain integrated measures, such as $\mathcal{L}^2$ and $|j|$, at much lower polynomial orders ($N_p \le 5$). However, accurate resolution of the Hartmann and Shercliff boundary layer dynamics, as measured through wall shear stresses, become the most stringent obstacle to numerical precision due to their large velocity gradients. Adequate grid independence was also seen for the quasi-two-dimensional model at $N_p=8$, and this component of the numerical scheme has also been previously validated in works by \citet{Ahmad:15} and \citet{Cassells:16}.
\begin{table}
  \begin{center}
\def~{\hphantom{0}}
  \begin{tabular}{lcccc}
      $N_p$  &   $\tau_{zx}\, (\times 10^{-1})$ & $\tau_{zy}\,(\times 10^{-3})$ & $\mathcal{L}^2$ & $|j|$\\[3pt]
       \,5  &~ 0.03207(1.0145) ~& 0.7508(3.947) ~& 2.9212(3.8154) ~& 0.04728(0.00238)\\
       \,6  &~ 0.03219(1.0164) ~& 0.7432(3.964) ~& 2.9212(3.8154) ~& 0.04728(0.00238)\\
       \,7  &~ 0.03229(1.0227) ~& 0.7463(3.975) ~& 2.9212(3.8154) ~& 0.04728(0.00238)\\
       \,8  &~ 0.03232(1.0247) ~& 0.7467(3.986) ~& 2.9212(3.8154) ~& 0.04728(0.00238)\\
       \,9  &~ 0.03233(1.0249) ~& 0.7469(3.987) ~& 2.9212(3.8154) ~& 0.04728(0.00238)\\
  \end{tabular}
  \caption{Grid sensitivity data for three-dimensional MHD base flow computations at $\Rey = 15\,000$ for two Hartmann number regimes; $\Ha=30$ and $\Ha=800$ (shown in the parentheses). The parameter $N_p$ indicates the polynomial order of the macro-element shape function. The wall shear stress inside a single Hartmann and Shercliff layer is indicated respectively by $\tau_{zx}$ and $\tau_{zy}$. The integrals of velocity magnitude $\mathcal{L}^2$ and current density $|j|$ over the flow domain are also given.}
  \label{tab:3d_base_conv}
  \end{center}
\end{table}

The accuracy of eigenmode predictions are dependent on the precision of the Krylov subspace iteration employed in the Arnoldi method, in addition to the resolution of the numerical methods. For the former, the iterative precision of the eigenvalue $\lambda$ and eigenvector $\hat{\BU}$ computations were measured by the residual
\BEQ
\label{residual}
r=\left\Vert{\mathscr{A}^\ast}{\mathscr{A}}\hat{\BU} - \lambda\hat{\BU}\right\Vert,
\EEQ
where $\Vert\bcdot\Vert$ is the standard vector norm. Eigenvalue convergence was deemed sufficient when $r < 10^{-7}$. For the latter, the grid independence of eigenvalue computations were ensured through a sufficiently high order polynomial degree shape function $N_p$. For Hartmann numbers $\Ha=30$ and $800$ at $\Rey =15\,000$, the optimal eigenvalue predictions as a function of $N_p$ for both three-dimensional and quasi-two-dimensional models are reported in table \ref{tab:3d_eig_conv}. For both Hartmann numbers, eigenvalue convergence of better than $0.002\%$ was achieved at $N_p = 8$; which is employed hereafter for all transient growth analyses.
\begin{table}
	\begin{center}
\def~{\hphantom{0}}
  \begin{tabular}{lcc}
      $N_p$  & ~~$\left\vert\lambda_{\text{opt}}\right\vert (\Ha = 30)~~$   &   $\left\vert\lambda_{\text{opt}}\right\vert(\Ha = 800)~~$ \\[3pt]
       ~5   & 464.894(17.218) & 4.907(4.295)\\
       ~6  & 457.221(16.921) &  5.168(4.937)\\
       ~7  & 466.184(17.703) &  5.080(4.626)\\
       ~8   & 467.273(17.721) & 5.061(4.829)\\
       ~9 & 467.214(17.720) & 5.059(4.831)\\
  \end{tabular}
  \caption{Grid sensitivity data for optimal eigenvalue computations using both three-dimensional and quasi-two-dimensional (shown in parentheses) MHD models at $\Rey = 15\,000$ for two Hartmann number regimes; $\Ha=30$ and $\Ha=800$. The parameter $N_p$ indicates the polynomial order of the macro-element shape function.}
  \label{tab:3d_eig_conv}
  \end{center}
\end{table}
For the three-dimensional eigenvalue computations, streamwise wavenumbers were investigated between $0\leq k \leq 80$, with the local maxima resolved to an accuracy of at least $0.1\%$ of the peak value. To ensure that the optimal growths were captured sufficiently, and a monotonic decay in amplifications were achieved at higher $\tau$, the analysis was conducted over time intervals extending to $\tau=5\TO$ for both three-dimensional and quasi-two-dimensional models. The optimal time intervals were resolved to at least 0.2 time units. The chosen numerical methodology and resolution were able to closely capture the same optimal growth rates predicted for hydrodynamic flow ($\Ha=0$), as well as the scaling in the range of $10 \leq \Ha \leq 50$ at $\Rey=5000$ obtained by \citet{Krasnov:10}. The linearised component of this solver has also been previously verified and implemented in works such as \citet{Hussam:12a}, \citet{Tsai:16} and \citet{Sapardi:17}.

\section{Linear transient growth}
\label{trans_grow}
\begin{figure}
\centering
\includegraphics[width=0.9\textwidth]{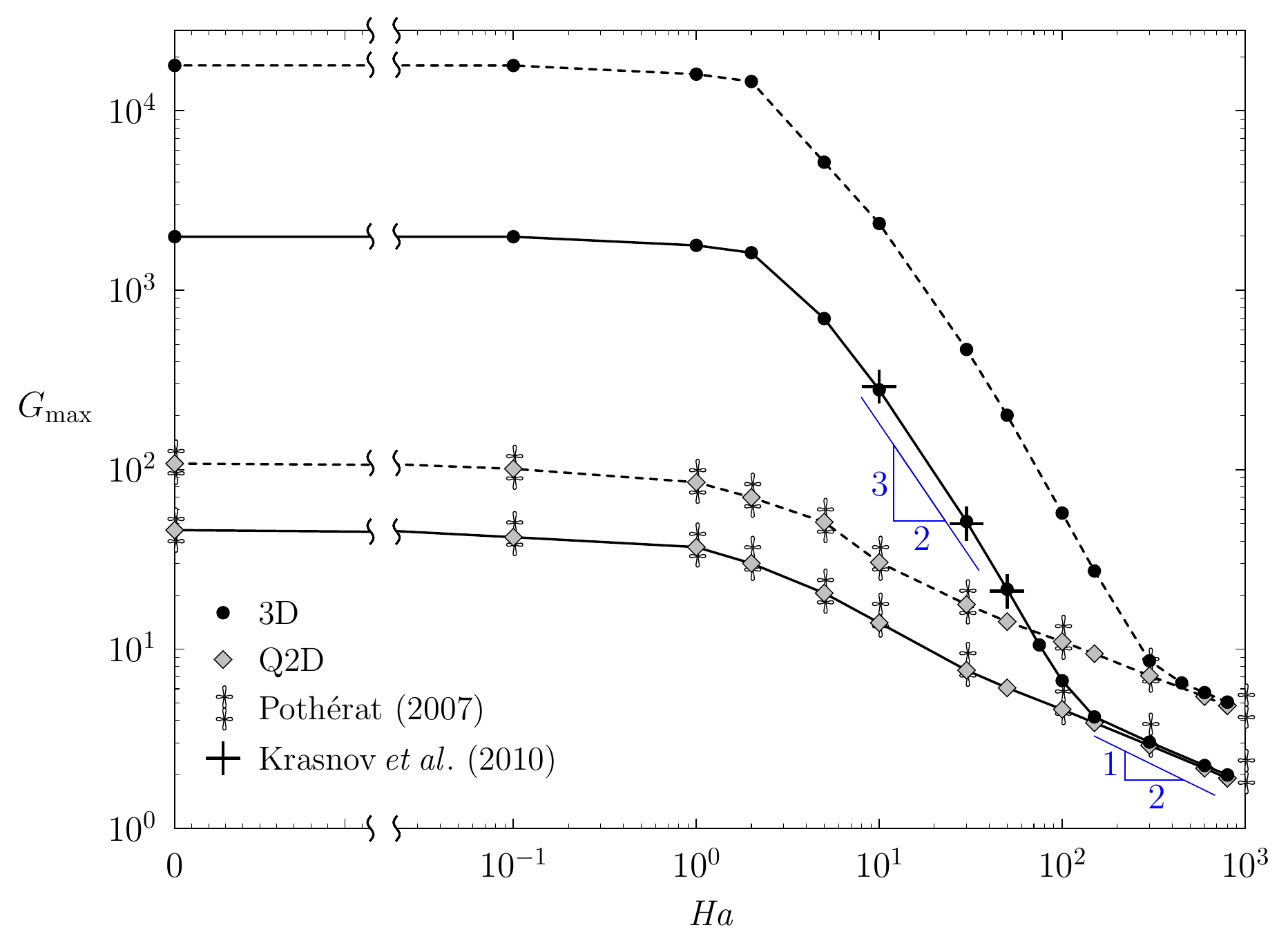}
\caption{Optimal transient growth as a function of Hartmann number for both three-dimensional and quasi-two-dimensional models at $\Rey=5000$ (\sampleline{}) and $15\,000$ (\sampleline{dashed}). Also provided are the optimal transient growths predicted by \protect\citet{Krasnov:10} using a three-dimensional analysis, and by \protect\citet{Potherat:07} using the SM82 model. The horizontal axis is logarithmically and linearly scaled for $\Ha \geq 2.2\times 10^{-2}$ and $\Ha < 2.2\times 10^{-2}$, respectively.}
\label{Gmax}
\end{figure}

The optimal transient growth and dynamics are presented and discussed in \S~\ref{3D_opt_disc}. An analysis of the optimal mode topology over evolution from hydrodynamic to high magnetic field regimes, inclusive of a comparison with SM82 predictions, is provided in \S~\ref{sec:mod_top}. An analysis of the underlying three-dimesnional and Q2D growth mechanisms in conjuction with the physical time-scales governing dynamics is presented in \S~\ref{sec:TG_mec}. The optimal transient growth for both MHD models is shown in figure~\ref{Gmax} for $0\leq \Ha \leq 800$ at $\Rey = 5000$ and $15\,000$. Included in this figure are the optimal transient growths calculated from a three-dimensional analysis by \citet{Krasnov:10} and from the SM82 model by \citet{Potherat:07}.
\subsection{Optimal transient growth}
\label{3D_opt_disc}
\begin{figure}
\centering
\begin{tabular}{l}
\begin{subfigure}[t]{0.45\textwidth}
\small (\textit{a})
\end{subfigure}\hfill
\begin{subfigure}[t]{0.48\textwidth}
\small (\textit{b})
\end{subfigure}\\
\begin{subfigure}{.49\textwidth}
  \centering
  \includegraphics[width=1\linewidth,valign=t]{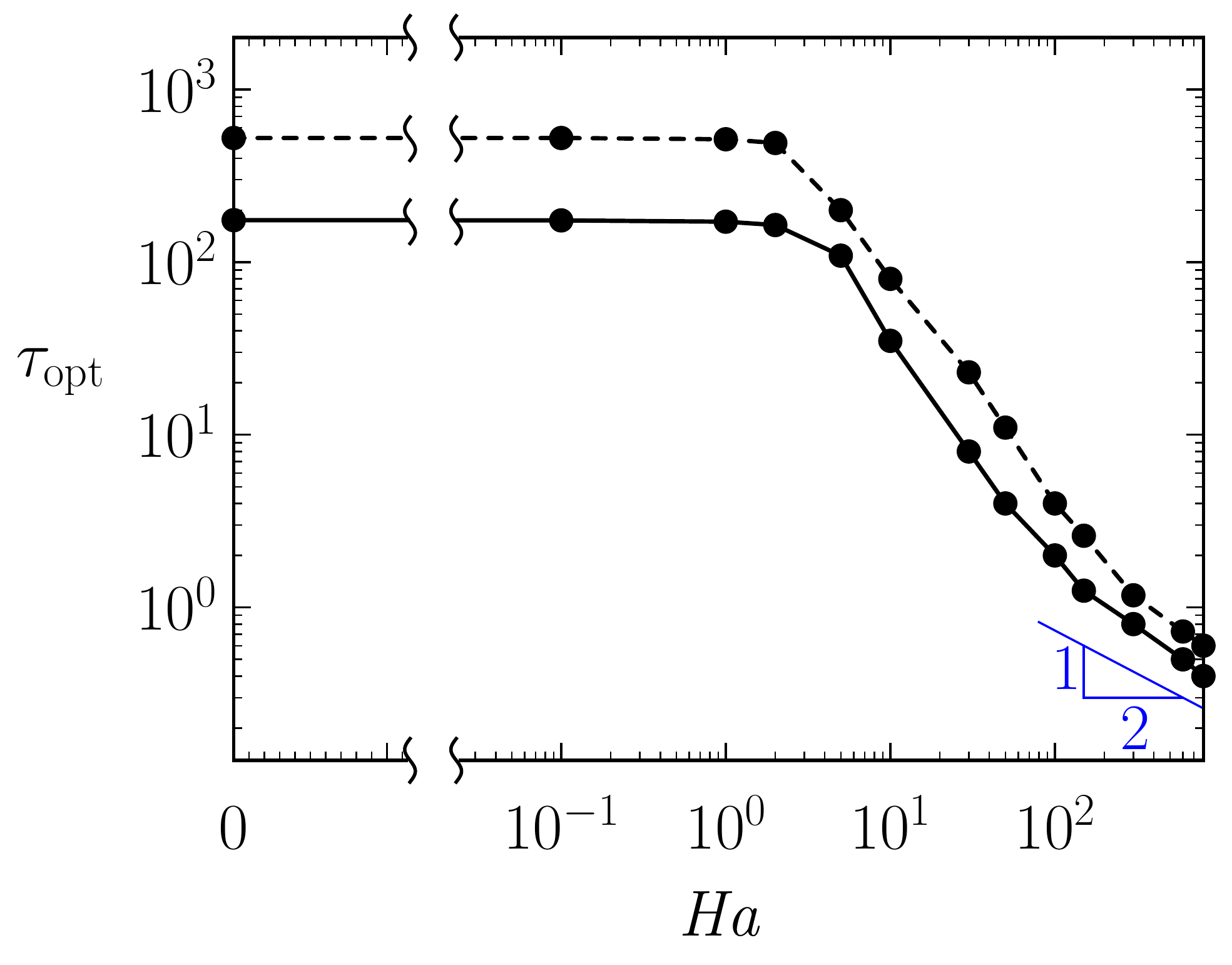}
\end{subfigure}%
\begin{subfigure}{.51\textwidth}
  \centering
\includegraphics[width=1\linewidth,valign=t]{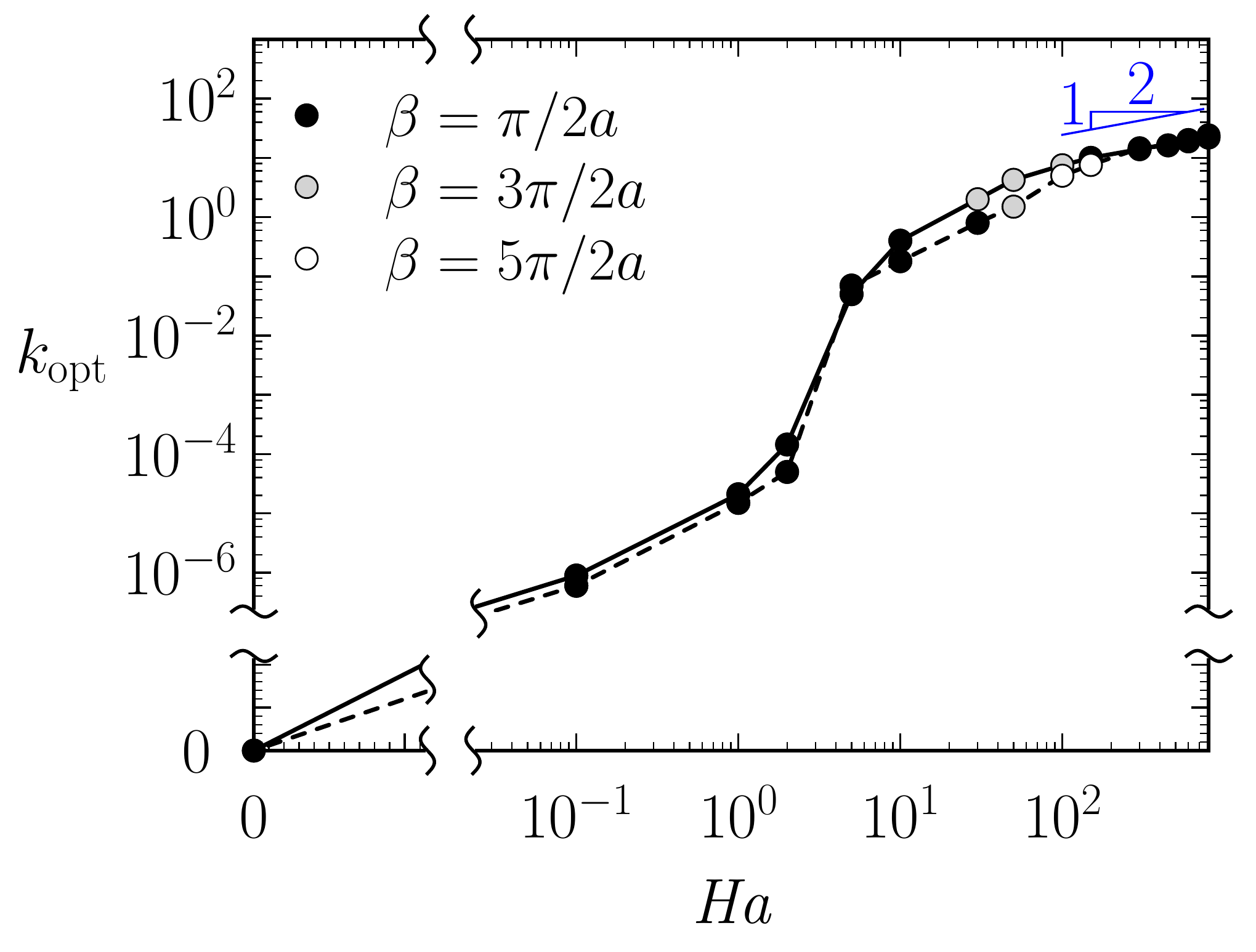}

\end{subfigure}
\end{tabular}
\caption{Optimal time interval (\textit{a}) and streamwise wavenumber (\textit{b}) as a function of Hartmann number for $\Rey = 5000$ (\sampleline{}) and $15\,000$(\sampleline{dashed}). The black, grey, and white symbols in (\textit{b}) represent the variation of the eigenvector periodicity in the magnetic field direction as a function of Hartmann number. The periodicity in this direction is quantified through the wavenumber $\beta = 2\pi/l_y$, where $l_y$ is the wavelength of the eigenvector in the vertical $y$-direction. The horizontal axis scaled as per figure~\ref{Gmax}. The vertical axis on (\textit{b}) is logarithmically and linearly scaled for $\KO \geq 2.2\times10^{-7}$ and $\KO < 2.2\times10^{-7}$, respectively.}
\label{fig:wav_tau}
\end{figure}

Transient growth is observed for all investigated cases. For three-dimensional analyses, the energy gain appears to be split into three broad regimes governed by distinctly different dynamics. For weak magnetic fields where the base velocity profile remains of a form consistent with Poiseuille duct flow (i.e. $\Ha \leq 1$), there is negligible change in attainable growth $\GM$, which remains to scale as $\sim\Rey^2$ largely independent of $\Ha$. Throughout this \textit{weak-MHD} regime the time interval for maximum growth $\TO$ is similarly $\Ha$-independent with scaling $\sim\Rey$. Conversely, a small, but finite, streamwise wavenumber dependence develops for non-zero field strengths, which remains mainly independent of $\Rey$ for all $\Ha$ cases investigated. Both of these respective behaviours are correspondingly depicted in figure \ref{fig:wav_tau}.

\begin{figure}
\centering
\begin{tabular}{l}
\begin{subfigure}[t]{0.45\textwidth}
\small (\textit{a})
\end{subfigure}\hfill
\begin{subfigure}[t]{0.48\textwidth}
\small (\textit{b})
\end{subfigure}\\
\begin{subfigure}{.5\textwidth}
  \centering
  \includegraphics[width=1\linewidth]{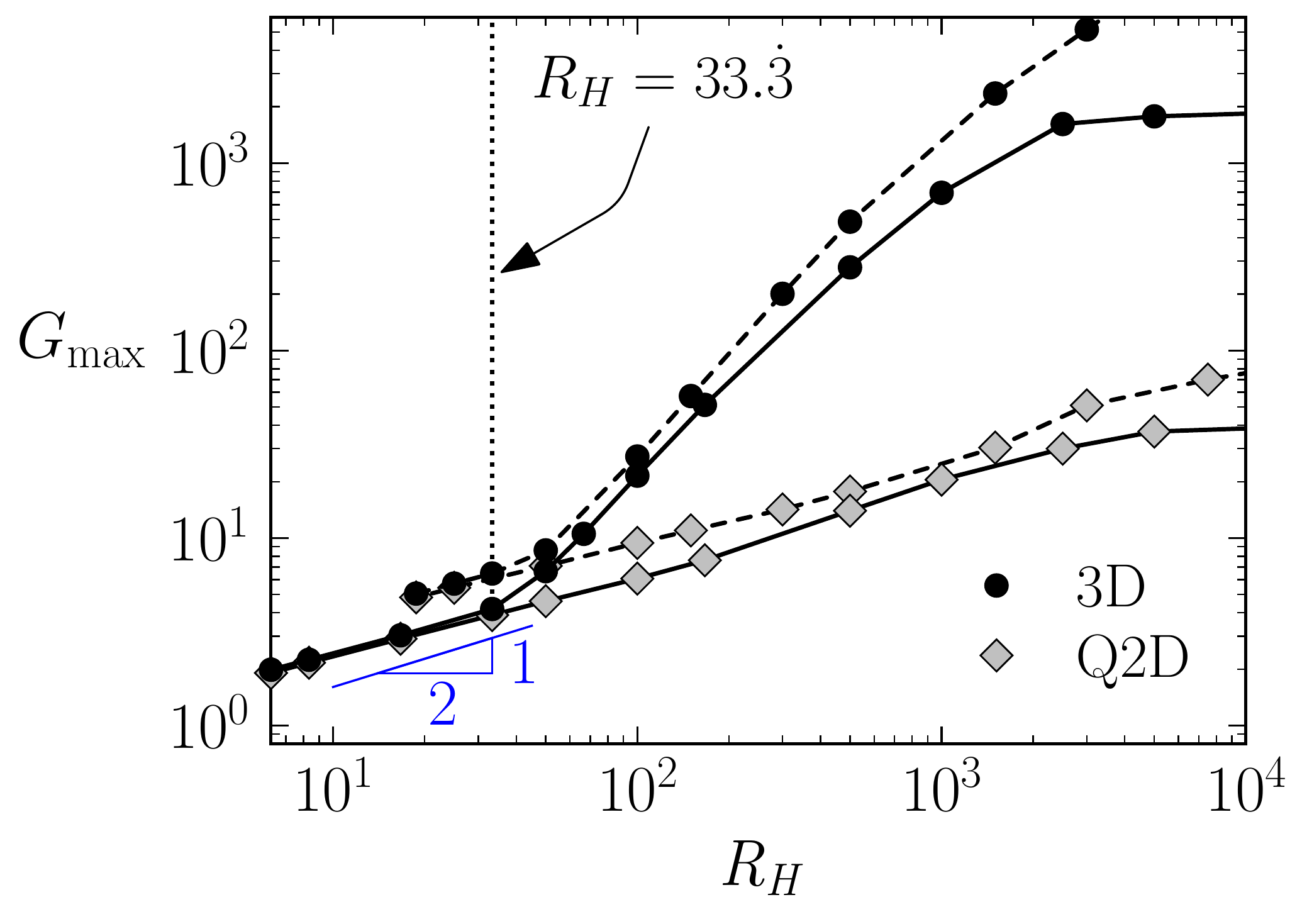}

\end{subfigure}%
\begin{subfigure}{.5\textwidth}
  \centering
 \includegraphics[width=1\linewidth]{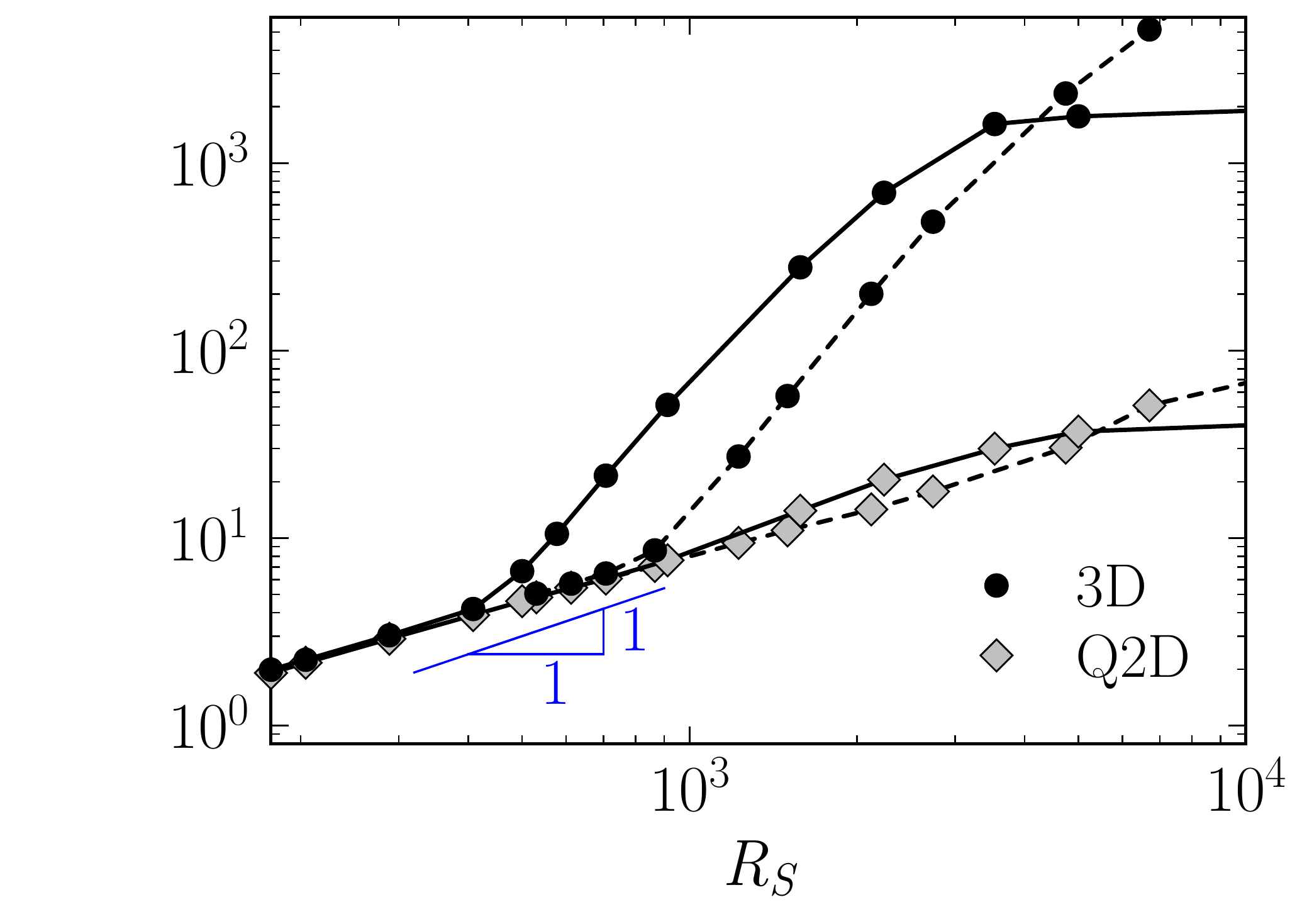}

\end{subfigure}
\end{tabular}
\caption{Optimal transient growth plotted against the Reynolds based on (\textit{a}) the Hartmann layer and (\textit{b}) Shercliff layer thickness $R_H$ and $R_S$, respectively, for both three-dimensional and quasi-two-dimensional models at $\Rey=5000$ (\sampleline{}) and $15\,000$ (\sampleline{dashed}).}
\label{Gmax_R_Rs}
\end{figure}

For $\Ha>1$, as the base flow topology morphs away from the hydrodynamic solution, the growth is strongly suppressed and correlated with a sharp decrease in the time horizon and streamwise wavelength for maximum amplification. Here the optimal time interval develops a strong dependence on $\Ha$ and is now relatively independent of $\Rey$. It is within this moderate-field-strength regime (hereafter \textit{moderate-MHD} regime) in which close agreement with \citet{Krasnov:10} in terms of the optimal growth $\GM$ and scaling of $\GM \propto \Ha^{-1.5}$ is seen for $10 \leq \Ha \leq 50$ at $\Rey=5000$. However, given the narrow Hartmann number range over which this relation holds, it is unlikely that this represents a true power law scaling. The streamwise wavenumber of optimal disturbances is found to depend only on the Hartmann number (similarly for $\Ha\lesssim 1$), indicating that eigenvector modulation is due primarily to electro-magnetic, rather than inertially driven, phenomena. 

A final asymptotic regime is reached when the Reynolds number at the scale of the Hartmann layer thickness $R_H=\Rey/\Ha$ reaches a transitional value of no lower than $R_H\approx 33.\dot{3}$ (this finding is supported with supplementary investigations at $\Rey = 2000$ and $10\,000$ which are not published here). Below this value, $\GM$ exclusively follows a relationship governed by the Reynolds number built upon the Shercliff layer thickness $R_S = \Rey/\Ha^{1/2}$. The respective regimes of governance of these parameters are readily illustrated in figures \ref{Gmax_R_Rs}(\textit{a}) and (\textit{b}) in which $\GM$ is plotted against $R_H$ and $R_S$ for a subset of the Hartmann numbers investigated. It is within this high Hartmann number regime where the flow is close to quasi-two-dimensional (hereafter \textit{Q2D-MHD} regime) and a remarkable agreement with SM82 model predictions are found. Here, not only is the scaling of energy growth $G_\text{max} \propto \Ha^{-0.44 \,\pm\, 0.07}$ closely captured by SM82, but also, strikingly, the amplification magnitudes are almost identical. The scaling of energy amplification is consistent with that known to dictate Q2D MHD critical linear and energy stability dynamics governed by Shercliff layer thickness $\delta_{S} = O(\Ha^{-1/2})$ \citep{Potherat:07, Vo:17}. The tendency of transient growth dynamics towards two-dimensionality is further illuminated by the transition of the optimal time interval and wavelength to a scaling of $\Ha^{-1/2}$ and $\Ha^{1/2}$, respectively. The mechanisms leading to these effects are discussed in more detail throughout \S~\ref{sec:mod_top} and \S~\ref{sec:TG_mec}.
\subsection{Optimal mode topology}
\label{sec:mod_top}
\begin{figure}
\centering
\begin{tabular}{l}
\\ \small (\textit{a}) $\Ha=0$, $R_H=\infty$, $R_S=\infty$ \hskip 65pt \small (\textit{b}) $\Ha=5$, $R_H=1000$, $R_S=2236.07$\\
\begin{subfigure}{.5\textwidth}
  \centering
  \includegraphics[width=1\linewidth]{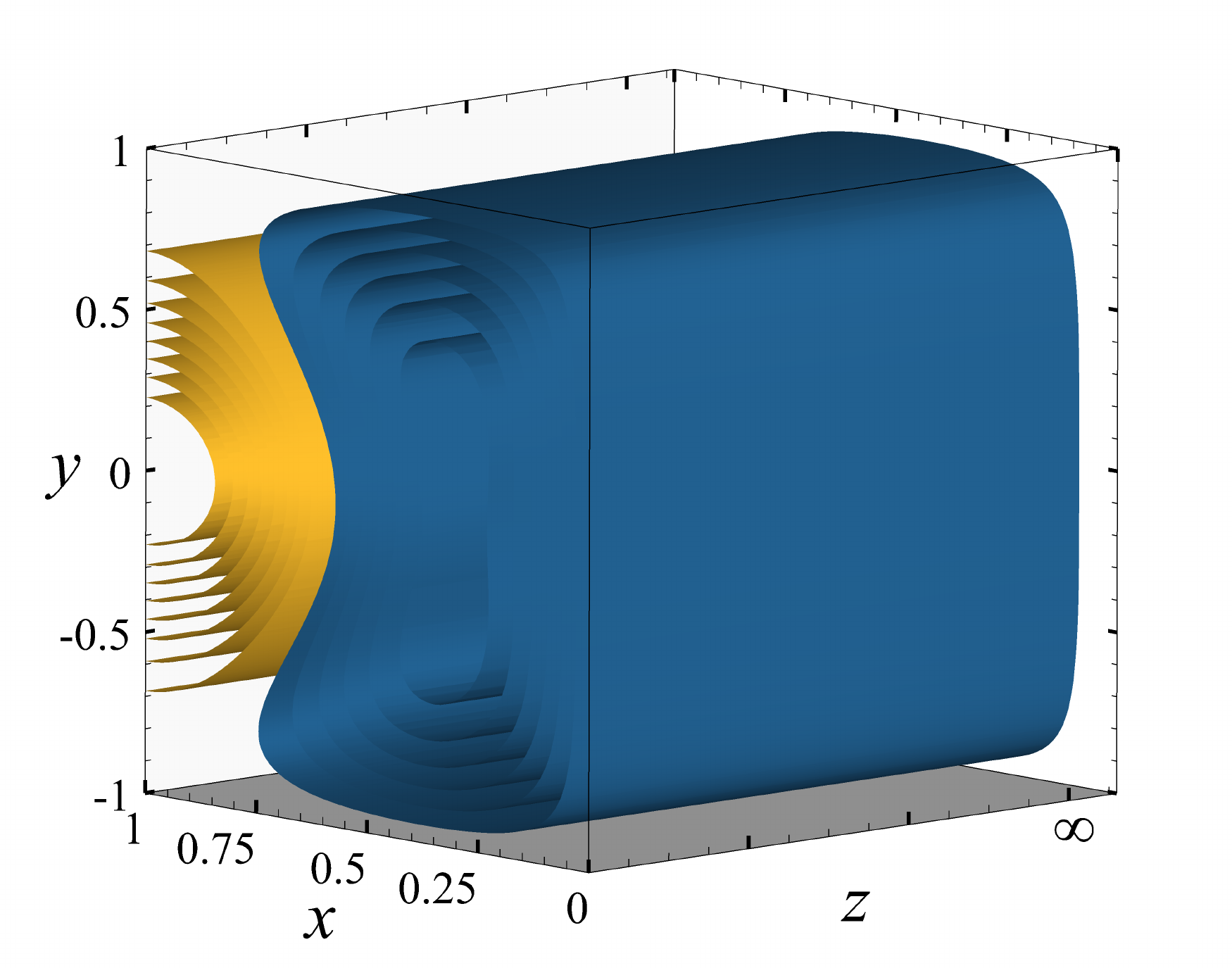}
\end{subfigure}
\begin{subfigure}{.5\textwidth}
  \centering
  \includegraphics[width=1\linewidth]{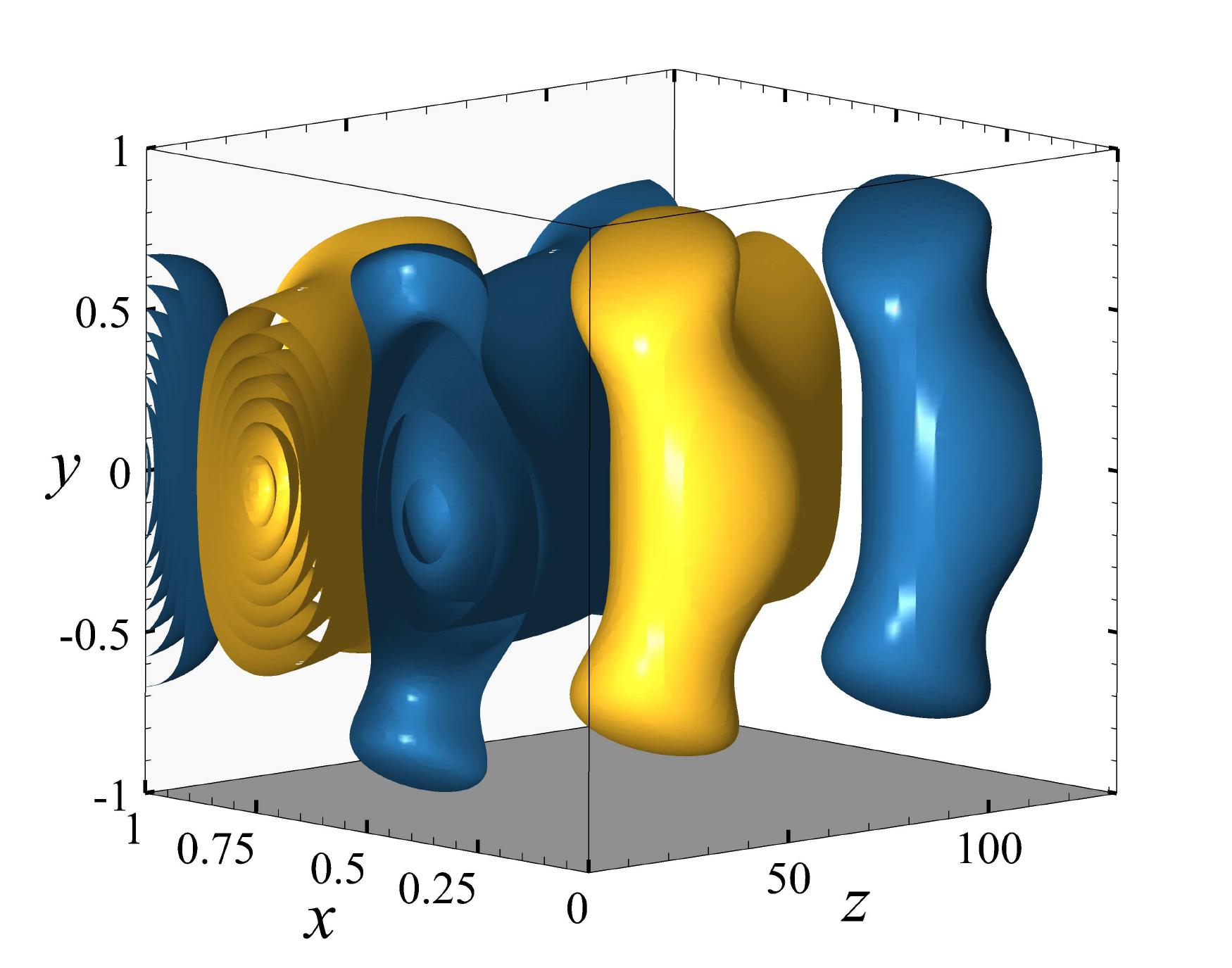}
\end{subfigure}
\\\\ \small (\textit{c}) $\Ha=10$, $R_H=500$, $R_S=1581.14$ \hskip 35pt \small (\textit{d}) $\Ha=50$, $R_H=100$, $R_S=707.11$\\
\begin{subfigure}{.5\textwidth}
  \centering
  \includegraphics[width=1\linewidth]{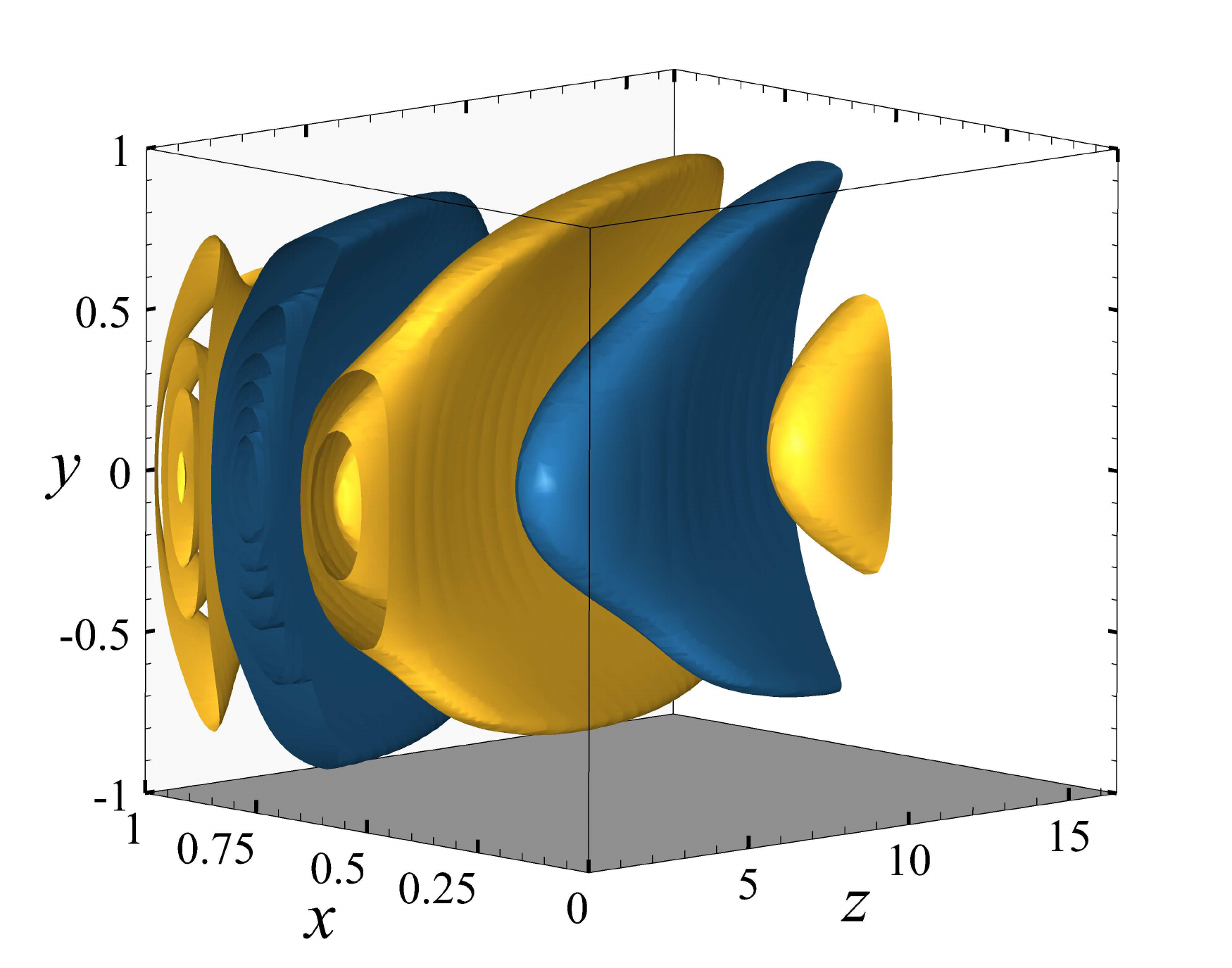}
\end{subfigure}
\begin{subfigure}{.5\textwidth}
  \centering
  \includegraphics[width=1\linewidth]{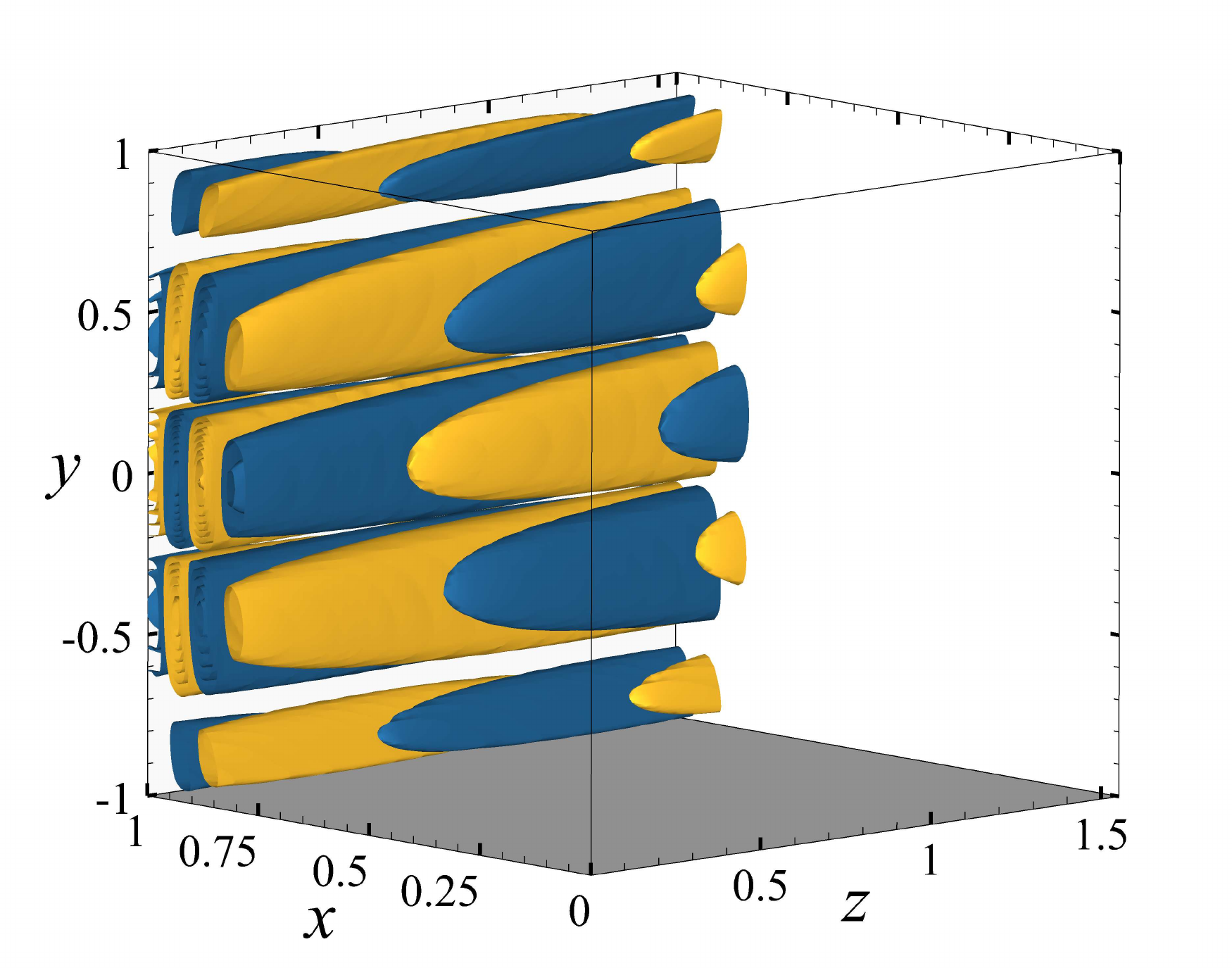}
\end{subfigure}
\\\\ \small (\textit{e}) $\Ha=150$, $R_H=33.\dot{3}$, $R_S=408.25$ \hskip 35pt \small (\textit{f}) $\Ha=300$, $R_H=16.\dot{6}$, $R_S=288.68$\\
\begin{subfigure}{.5\textwidth}
  \centering
 \includegraphics[width=1\linewidth]{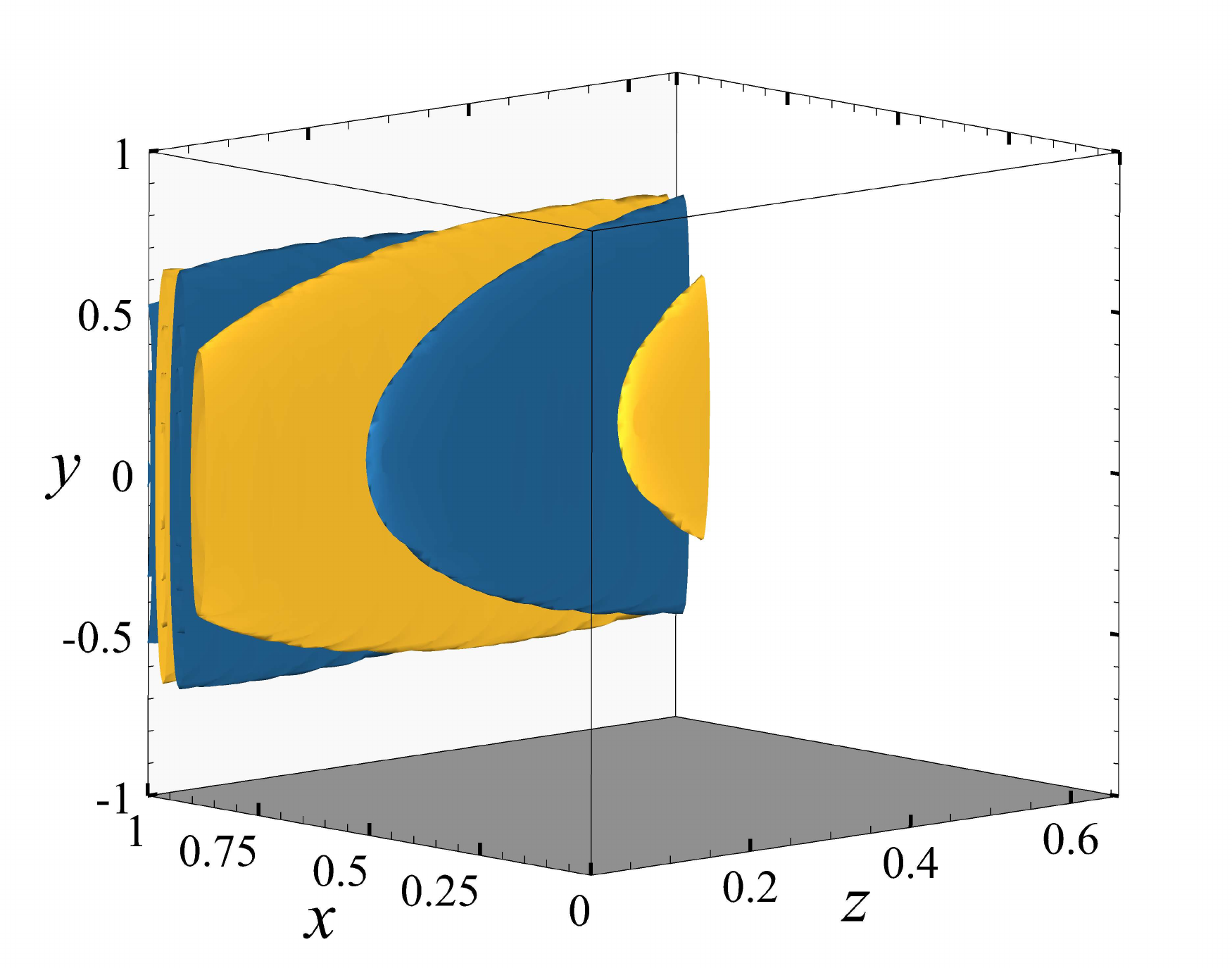}
\end{subfigure}
\begin{subfigure}{0.5\textwidth}
  \centering
 \includegraphics[width=1\linewidth]{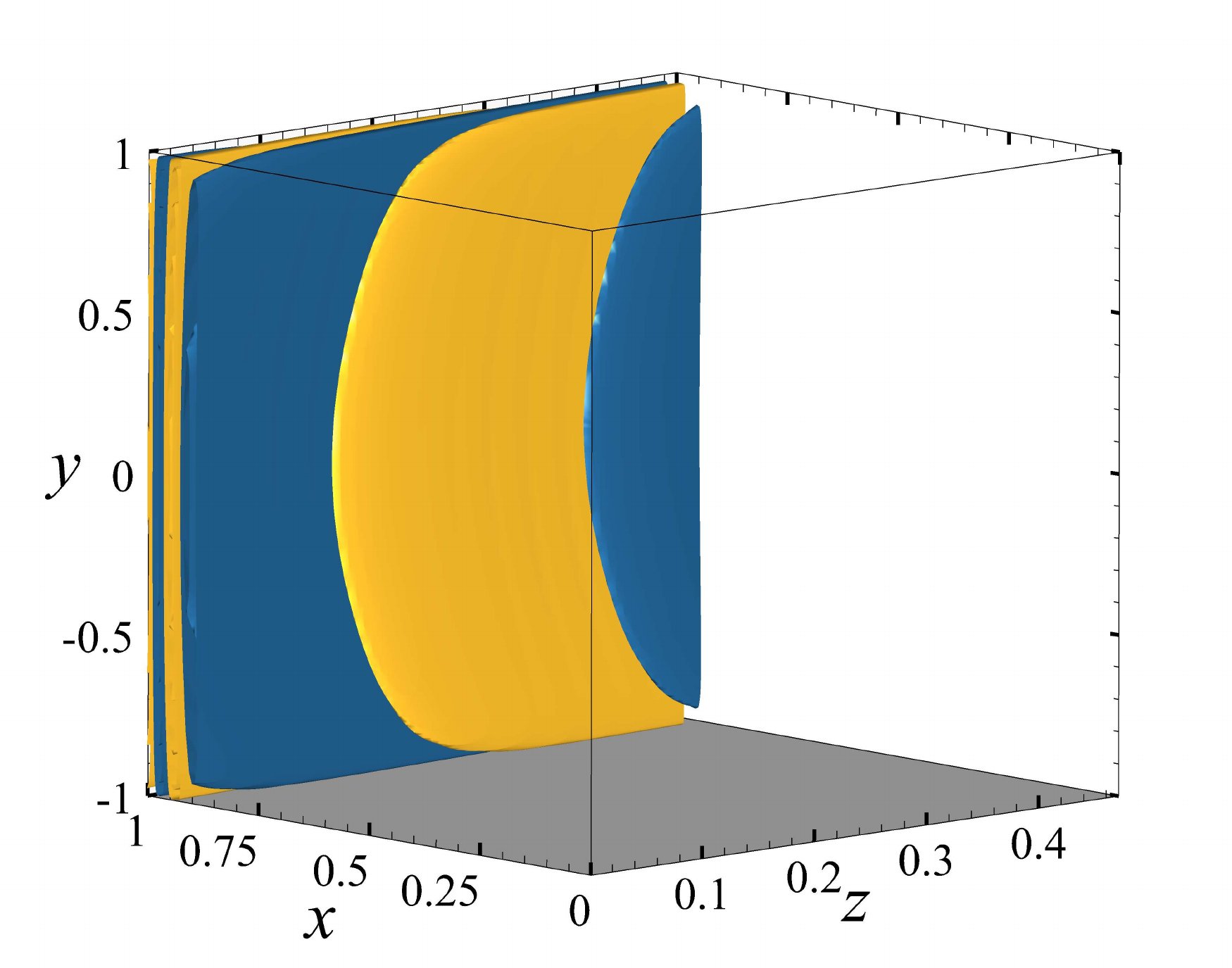}
\end{subfigure}
\end{tabular}
\caption{Vertical component of vorticity $\omega_{y}$ iso-surfaces of the optimal eigenvector fields producing maximum transient amplification $G_{\text{max}}$ for (\textit{a}--\textit{f}) $\Ha=0$, $5$, $10$, $50$, $150$ and $300$ at $\Rey=5000$. Blue and yellow contours represent positive and negative vorticity, respectively. Contour levels are adjusted to approximately 90\% of the maximum magnitude of $\omega_{y}$. The flow is from left to right in the positive $z$-direction, with the magnetic field orientated vertically in the positive $y$-direction. For clarity, the iso-surfaces are only plotted for $0 \leq x \leq 1$.}
\label{eig_xz_vort_Re5k}
\end{figure}
\begin{figure}
\centering
\begin{tabular}{l}
\\ \small (\textit{a}) $\Ha=50$, $R_H=300$, $R_S=2121.32$ \hskip 35pt \small (\textit{b}) $\Ha=150$, $R_H=100$, $R_S=1224.74$\\
\begin{subfigure}{.5\textwidth}
  \centering
  \includegraphics[width=1\linewidth]{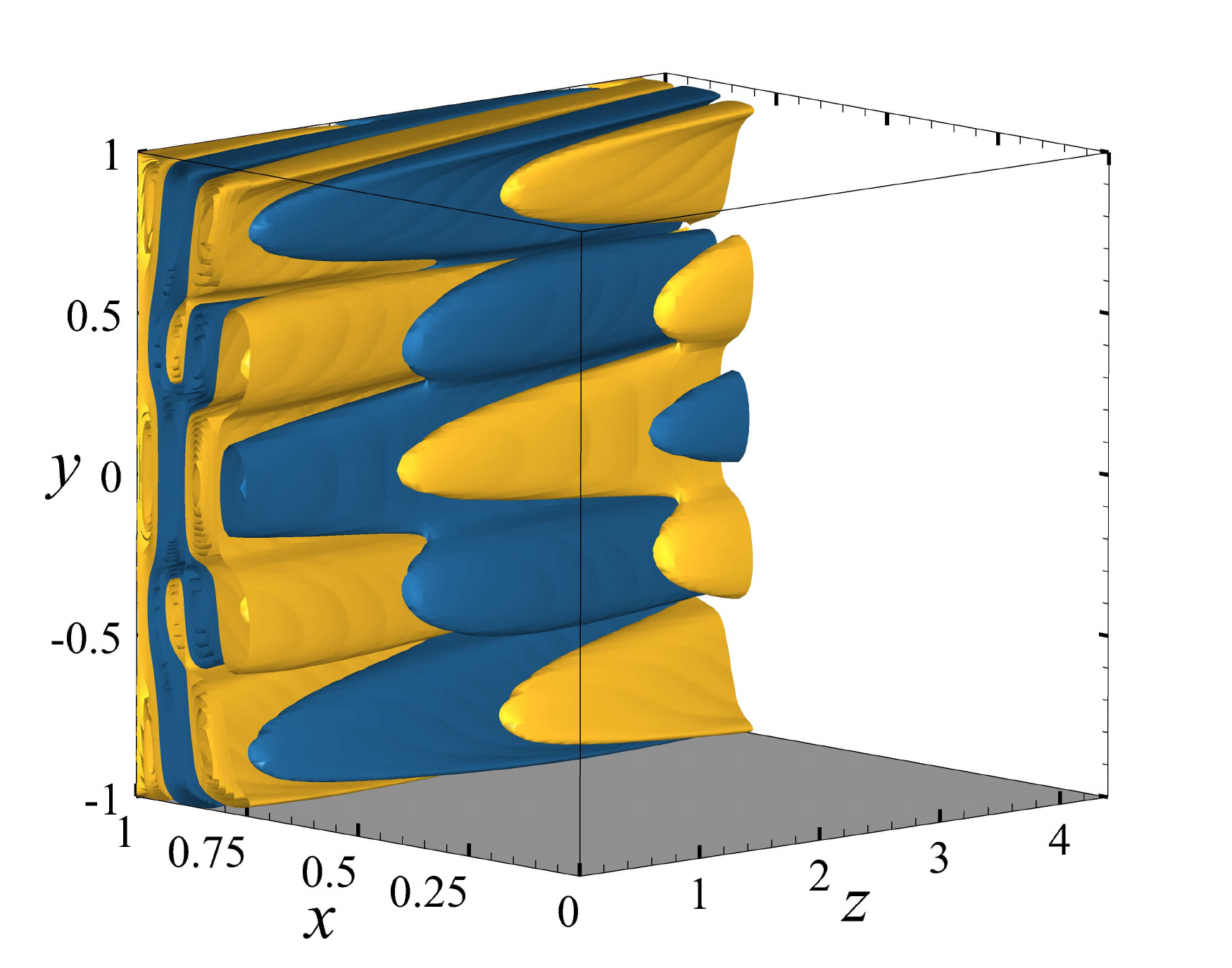}
\end{subfigure}
\begin{subfigure}{.5\textwidth}
  \centering
  \includegraphics[width=1\linewidth]{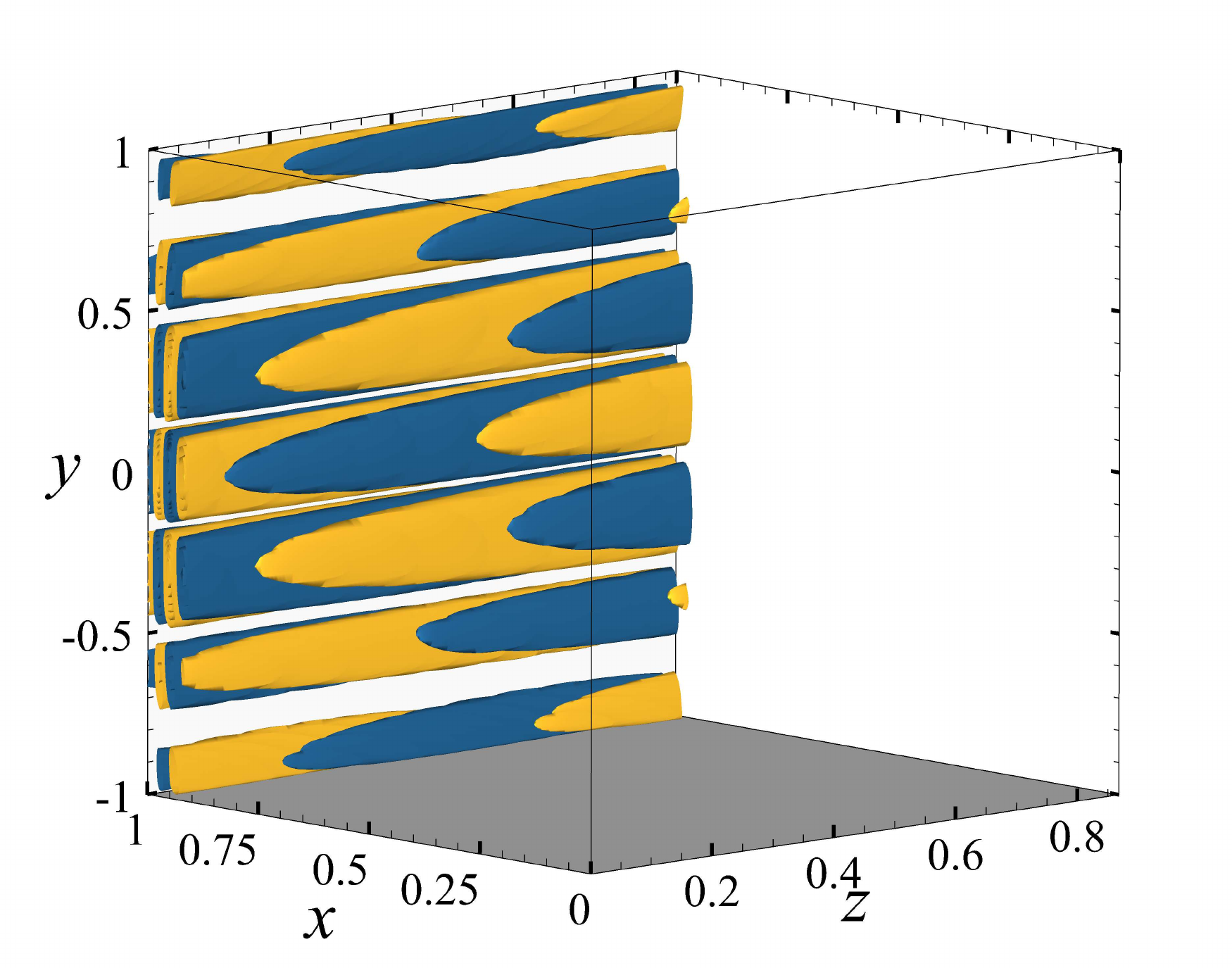}
\end{subfigure}
\\\\ \small (\textit{c}) $\Ha=300$, $R_H=50$, $R_S=866.03$ \hskip 40pt \small (\textit{d}) $\Ha=800$, $R_H=18.75$, $R_S=530.33$\\
\begin{subfigure}{.5\textwidth}
  \centering
 \includegraphics[width=1\linewidth]{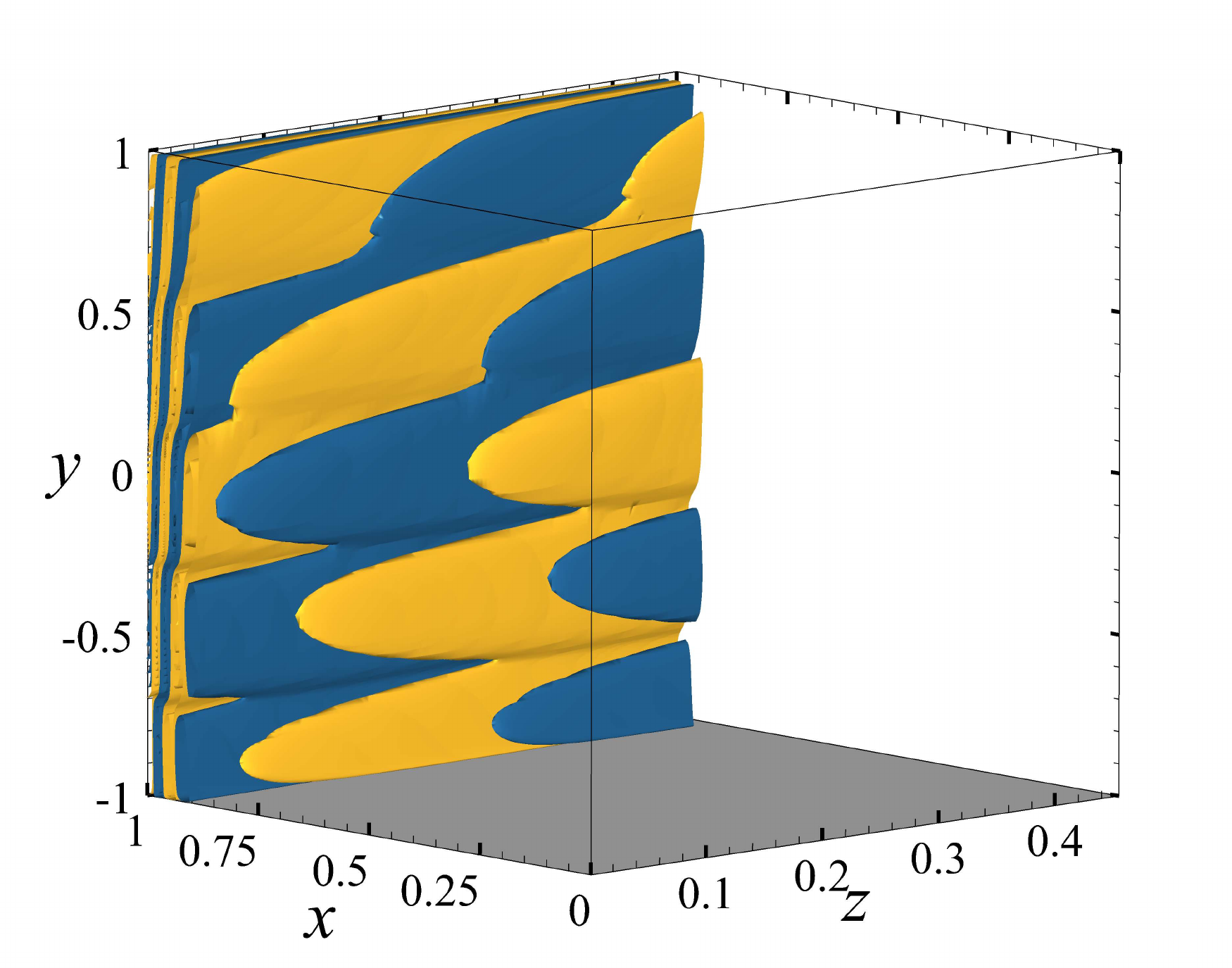}
\end{subfigure}
\begin{subfigure}{0.5\textwidth}
  \centering
 \includegraphics[width=1\linewidth]{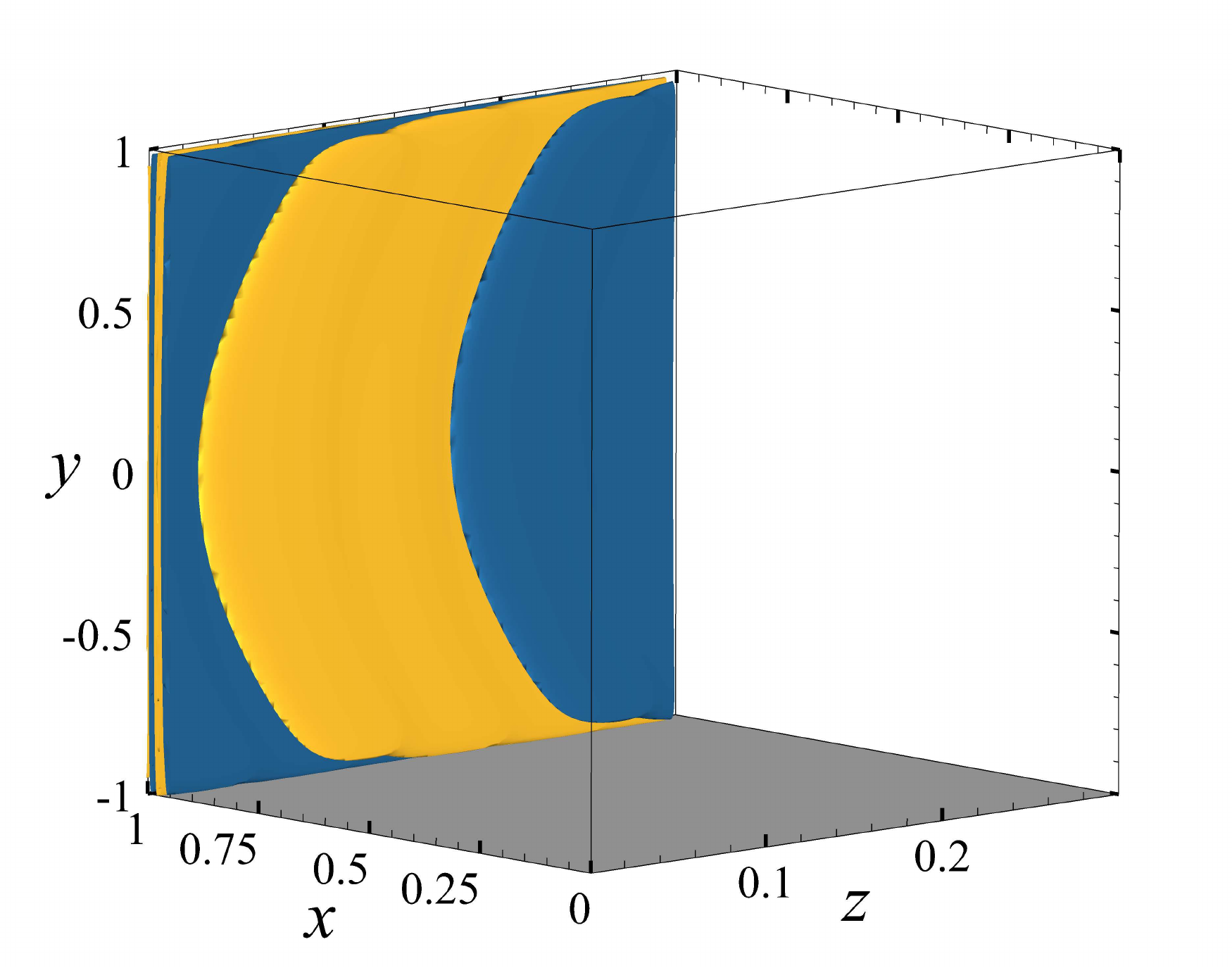}
\end{subfigure}
\end{tabular}
\caption{Vertical component of vorticity $\omega_{y}$ iso-surfaces of the optimal eigenvector fields producing maximum transient amplification $G_{\text{max}}$ for (\textit{a}--\textit{d}) $\Ha=50$, $150$, $300$ and $800$ at $\Rey=15\,000$. Iso-surfaces specified as per figure~\protect\ref{eig_xz_vort_Re5k}.}
\label{eig_xz_vort_Re15k}
\end{figure}
\begin{figure}
\centering
\begin{tabular}{l}
\begin{subfigure}[t]{0.45\textwidth}
\small (\textit{a})
\end{subfigure}\hfill
\begin{subfigure}[t]{0.48\textwidth}
\small (\textit{b})
\end{subfigure}\\
\begin{subfigure}{.5\textwidth}
  \centering
  \includegraphics[width=1\linewidth]{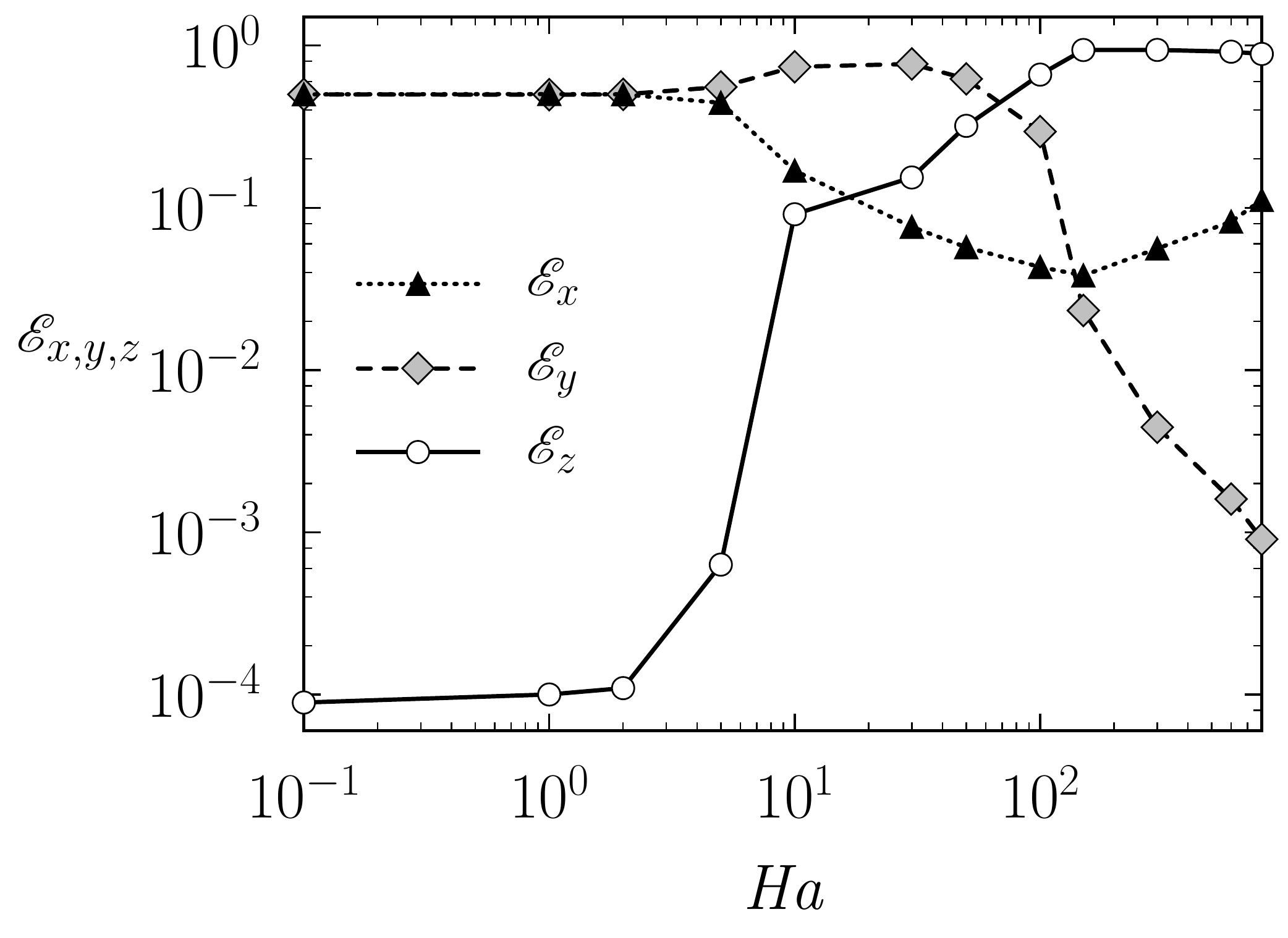}

\end{subfigure}%
\begin{subfigure}{.5\textwidth}
  \centering
 \includegraphics[width=1\linewidth]{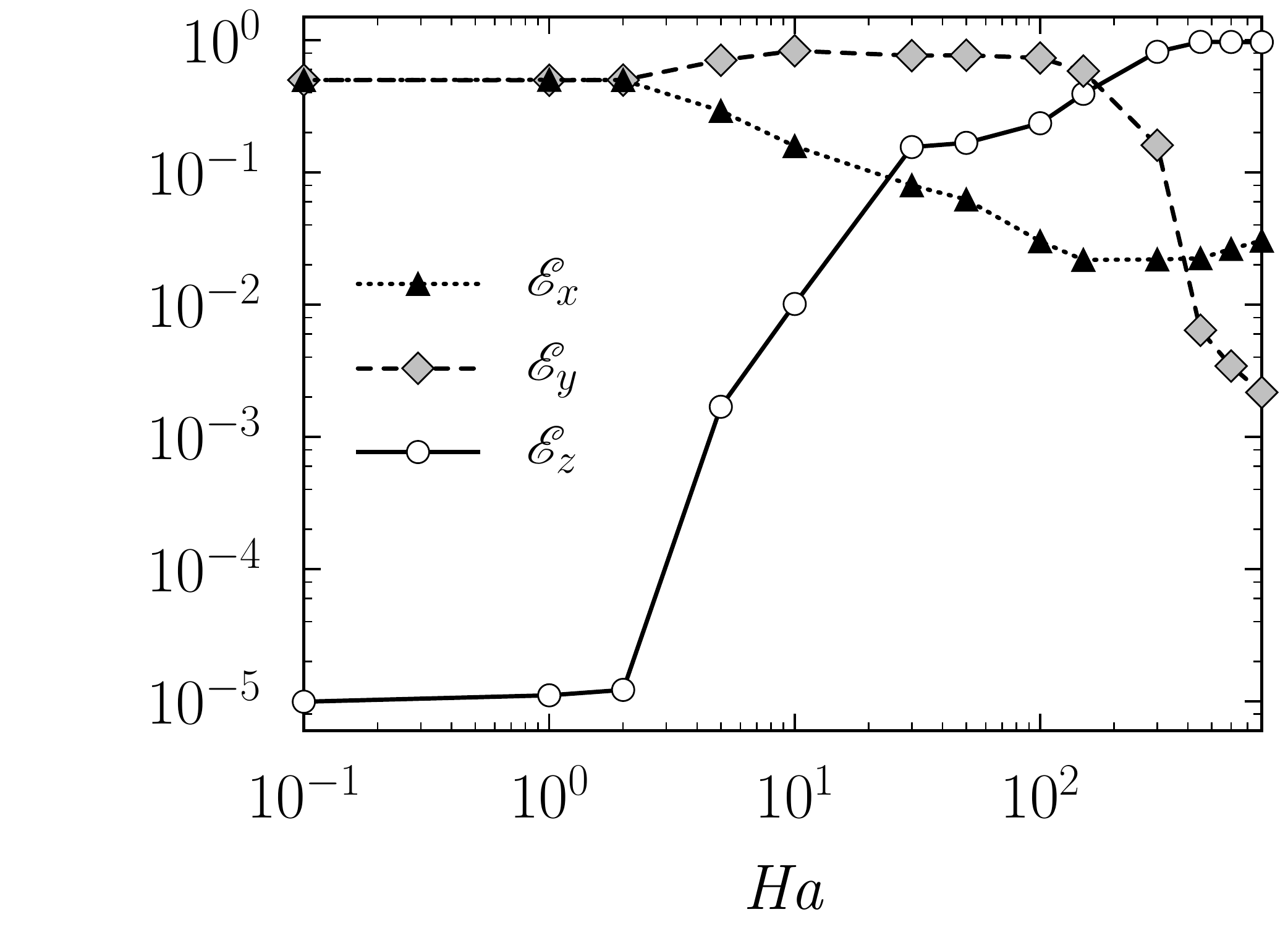}

\end{subfigure}
\end{tabular}
\caption{Distribution of initial eigenvector energy across horizontal $\mathscr{E}_x$, vertical $\mathscr{E}_y$ and streamwise $\mathscr{E}_z$ velocity components (i.e. $u,v$ and $w$), as a function of $\Ha$ for (\textit{a}) $\Rey =5000$ and  (\textit{b}) $\Rey =15 \,000$. Energy magnitudes are normalised using the total perturbation kinetic energy.}
\label{fig:vel_en}
\end{figure}

In this section the optimal eigenvectors are analysed in terms of their spatial form and initial componental energy distribution. Attention is focused on the three-dimensional optimal modes and how their characteristics change over transition from weak-MHD to Q2D-MHD regimes. In addition, a comparison between three-dimensional and SM82 optimal mode topology upon evolution to $\TO$ in the Q2D-MHD regime is presented. The initial disturbance topologies leading to optimal transient growth are visualised as iso-surfaces of the vertical component of vorticity $\omega_{y}$ in figure~\ref{eig_xz_vort_Re5k} for $0 \leq \Ha \leq 300$ at $Re=5000$ and figure~\ref{eig_xz_vort_Re15k} for $50 \leq \Ha \leq 800$ at $\Rey=15\,000$. Displayed in figure~\ref{fig:vel_en} is the distribution of the total initial kinetic energy across the horizontal $\mathscr{E}_x$, vertical $\mathscr{E}_y$ and streamwise $\mathscr{E}_z$ velocity components (i.e. $u,v$ and $w$). 

In the absence of a magnetic field, the optimal eigenvectors  present as streamwise independent and aligned counter-rotating vortices, as shown in figure~\ref{eig_xz_vort_Re5k}(\textit{a}). For weak magnetic field strengths $0<\Ha \lesssim 1$, the only qualitative change in topology is that of slight wavelength dependence (i.e. $\KO\rightarrow 0$ as $\Ha\rightarrow 0$). Thus, the initial energy in the streamwise velocity component is substantially smaller compared to the spanwise components, and the modes retain close similarity to the algebraic instabilities producing maximum transient amplification in wall-bounded hydrodynamic shear flows \citep{Biau:08, Schmid:01}. In contrast, optimal mode topology in the moderate-MHD regime appears to be governed by $R_H$. Upon transition to this regime, the optimal modes form obliquely-aligned streamwise rolls overlapping within the sidewall boundary layers and slanted out from the wall in the upstream direction, which is is readily highlighted in figures~\ref{eig_xz_vort_Re5k}(\textit{b},~\textit{c}). Streamwise independence is no longer present, and as seen in figure~\ref{fig:vel_en}, a substantial increase in  $\mathscr{E}_z$ relative to $\mathscr{E}_x$ and $\mathscr{E}_y$ is observed. Interestingly for both $\Rey$ at $\Ha\gtrsim 2$,  $\mathscr{E}_x$ and $\mathscr{E}_y$ start to deviate away from one another due to a relative increase and decrease in magnitude, respectively. By $\Ha \approx 30$, the energy contained in the streamwise component is larger than the horizontal constituent for both Reynolds numbers. 

It is within the moderate MHD-regime (i.e. $ 350\gtrsim R_H \gtrsim 33.\dot{3}$) where $\mathscr{E}_z$ becomes globally the largest component, albeit only at relatively large field strengths of $\Ha \approx 100$ and $\Ha\approx 300$ for $\Rey=5000$ and $15\,000$, respectively. The eigenvectors also present again as streamwise-aligned oblique modes in this regime, yet now with increased vertical periodicity. This behaviour is depicted in figures~\ref{eig_xz_vort_Re5k}(\textit{d})~and~\ref{eig_xz_vort_Re15k}(\textit{b},~\textit{c}), along with a quantification presented in figure~\ref{fig:wav_tau}(\textit{b}). This change in vertical wavelength represents a counter-intuitive modification to the largest field-parallel length scale of the disturbance structures $\ell_y$. The corresponding decrease in $\ell_y$ with increasing $\Ha$ in this parameter space is in stark contrast to the natural progression of eddy anisotropy which is often expected as field strengths are increased. Indeed anisotropy is caused by the Lorentz force increasing momentum diffusion of structures perpendicular to the field and the corresponding diffusivity increases linearly with $\bm{B_0}$ \citep{Sommeria:82,Potherat:00,Potherat:14}.  

However, upon transition to the final asymptotic regime $R_H \lesssim 33.\dot{3}$, a break-down in the periodicity of vortices in the vertical direction occurs, and a recovery of the parallel length scale tending towards that of the duct height $\ell_y \sim 2a$ is seen. As illustrated through figures \ref{eig_xz_vort_Re5k}(\textit{e},~\textit{f}) and \ref{eig_xz_vort_Re15k}(\textit{d}), when the relative field strength is increased, $\ell_y$ correspondingly grows so that it exceeds that of the duct height (i.e. $\ell_y \gtrsim 2a$); resulting in the eigenvector becoming nearly two-dimensional. Quantification of this behaviour is seen in the sharp decline of $\mathscr{E}_y$ in this regime due to the Lorentz force diffusing velocity differentials perpendicular to the field direction. Further to this, transition to this regime sees a change in the hierarchy of spanwise energy components such that $\mathscr{E}_x$ now dominates. The dominance of both the initial streamwise and horizontal energies over that of $\mathscr{E}_y$ is an additional signal to the transition to two-dimensionality of the optimal modes.

\begin{figure}
\centering
\begin{tabular}{l}
\\ \small (\textit{a}) $\Ha=800$, $\Rey=5000$, $R_H=6.25$, $R_S =176.78$\\
\begin{subfigure}{.5\textwidth}
  \centering
  \includegraphics[width=1\linewidth]{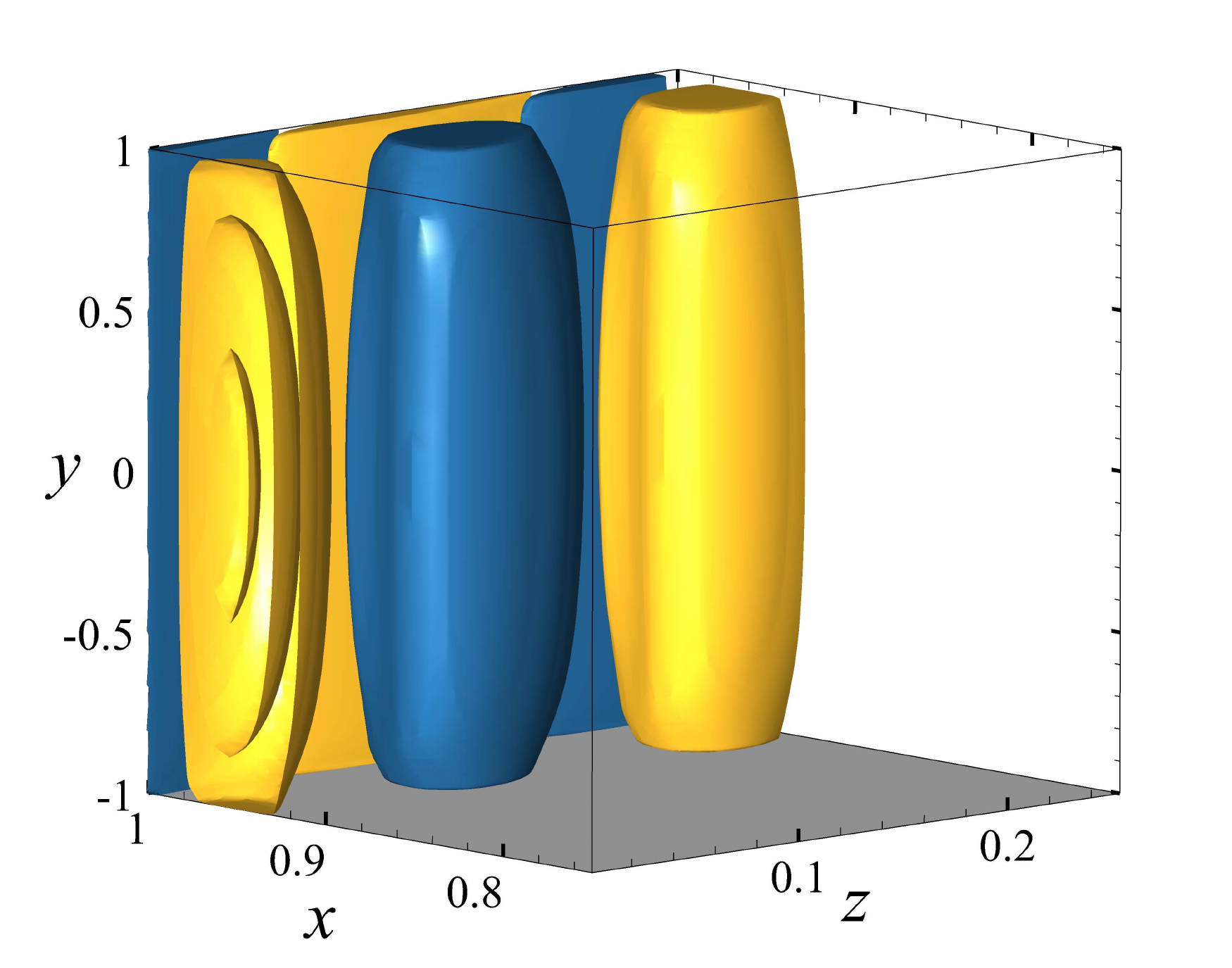}
\end{subfigure}
\begin{subfigure}{.5\textwidth}
  \centering
  \includegraphics[width=1\linewidth]{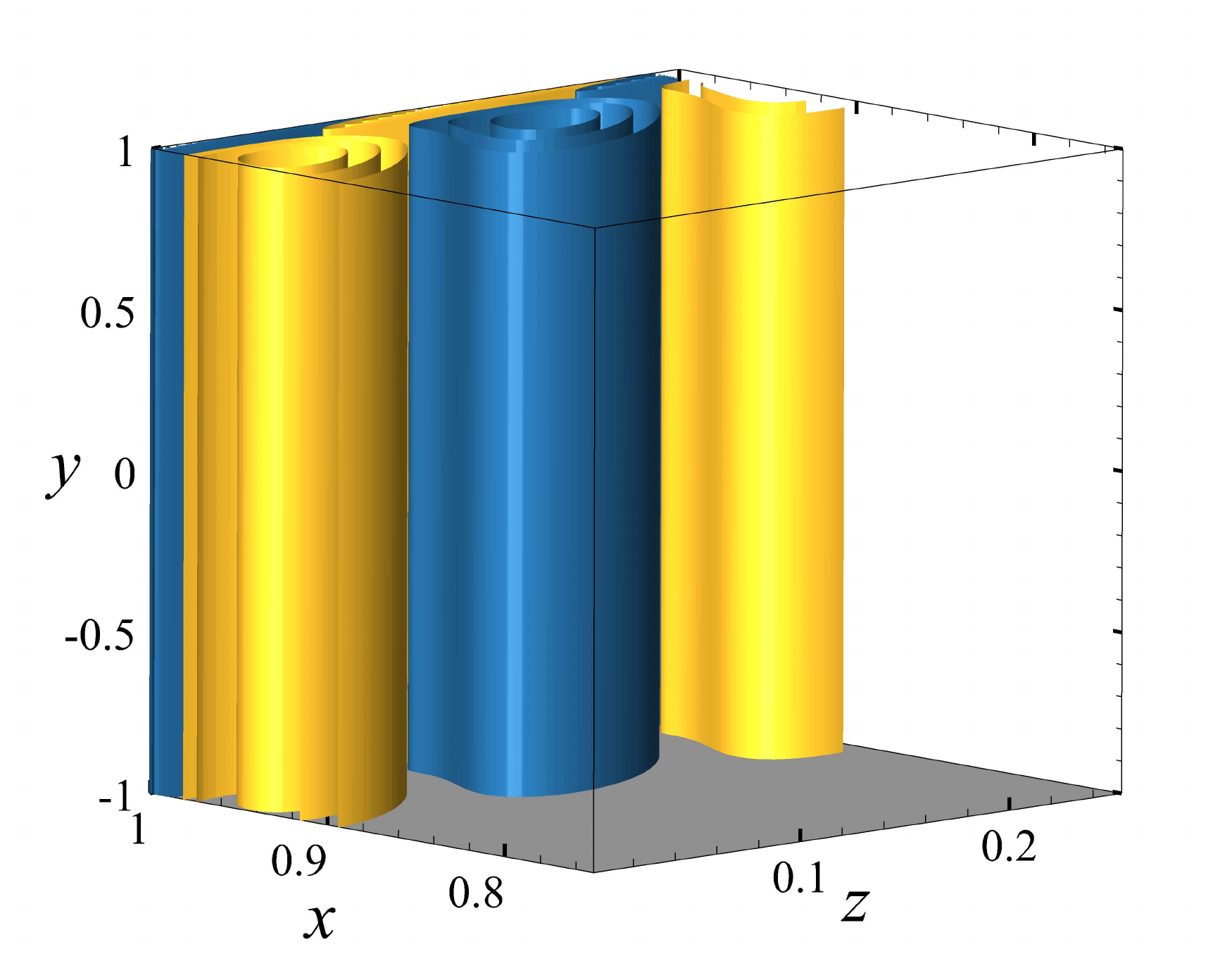}
\end{subfigure}
\\\\ \small (\textit{b}) $\Ha=800$, $\Rey=15\,000$, $R_H=18.75$, $R_S=530.33$\\
\begin{subfigure}{.5\textwidth}
  \centering
 \includegraphics[width=1\linewidth]{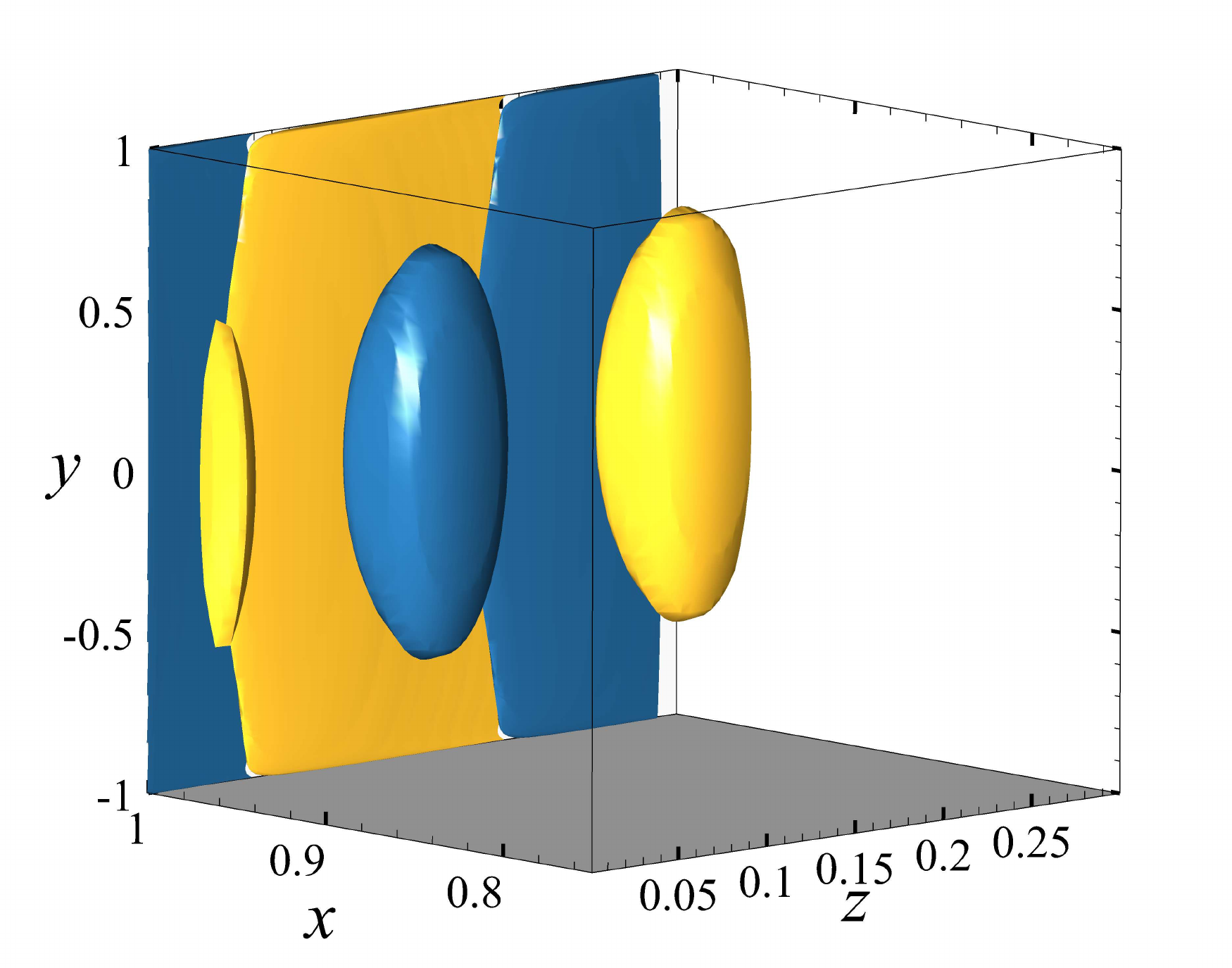}
\end{subfigure}
\begin{subfigure}{0.5\textwidth}
  \centering
 \includegraphics[width=1\linewidth]{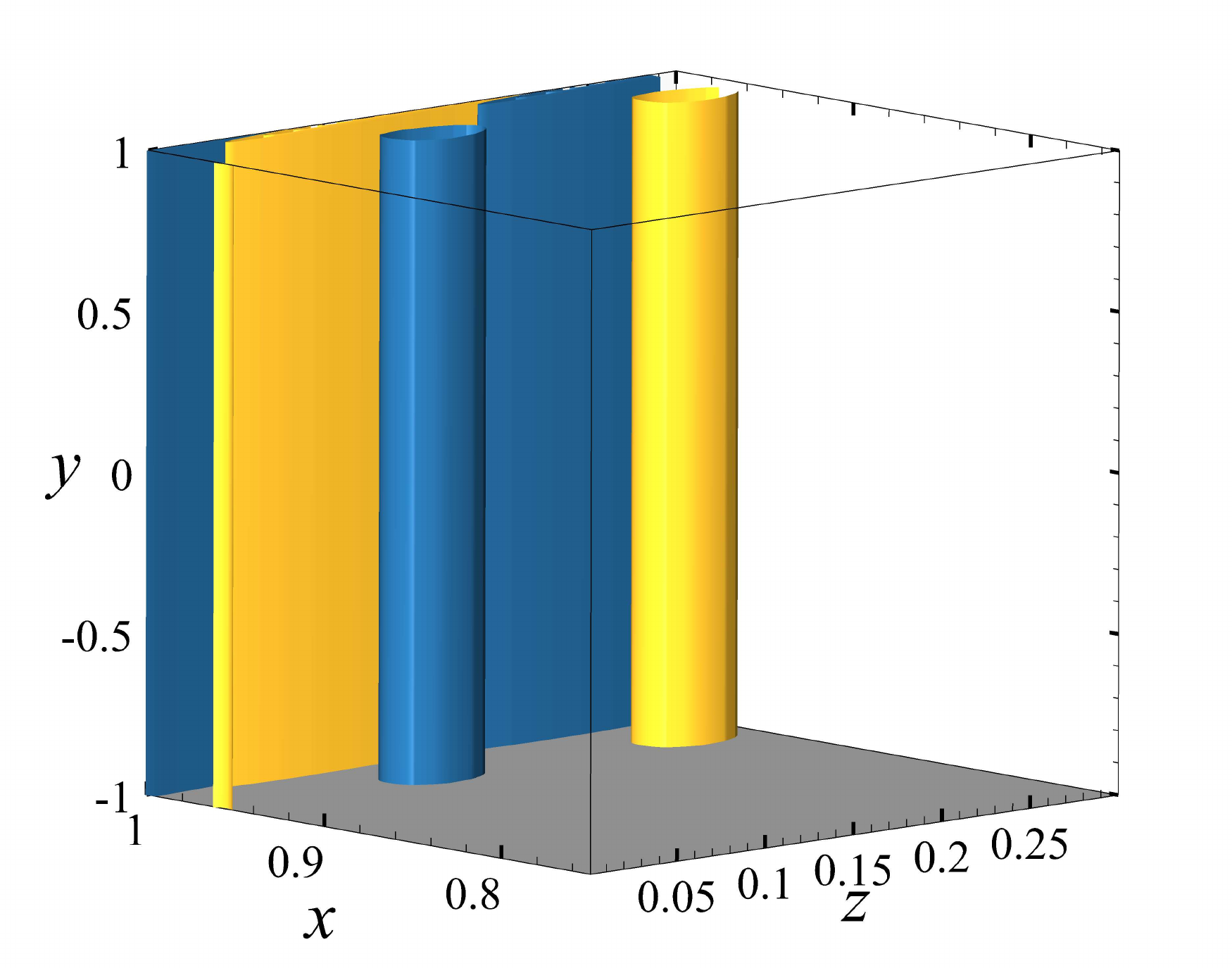}
\end{subfigure}
\end{tabular}
\caption{Vertical component of vorticity $\omega_{y}$ iso-surfaces of the perturbation fields at time of maximum transient amplification $\TO$ for $\Ha=800$ at (\textit{a}) $\Rey=5000$ and (\textit{b})~$\Rey=15\,000$. The leftmost figure in each of (\textit{a}) and (\textit{b}) are the iso-surfaces for three-dimensional analysis, the rightmost figures are a uniform extrusion of the $x$-$z$ planar Q2D optimal disturbance fields in $y$-direction. Iso-surface colours as per figure~\protect\ref{eig_xz_vort_Re5k}. For clarity, the iso-surfaces are only shown for $0.75 \leq x \leq 1$.}
\label{topt_q2d_v_3d}
\end{figure}

A primary aim of the present work is to determine the validity of SM82 in reproducing three-dimensional transient growth predictions. As such, the disturbance topology in the Q2D-MHD regime using both MHD approaches at time of maximum energy gain $\TO$ is presented for comparison in figure~\ref{topt_q2d_v_3d} for $\Ha=800$ at (\textit{a}) $\Rey=5000$ and (\textit{b}) $\Rey = 15\,000$. Here a uniform vertical extrusion of the SM82 disturbance fields is provided to help indicate the resultant averaging of the three-dimensional solution in the $y$-direction as per \eqref{SM82_lin}. Consistent with the excellent agreement of optimal energy growth between both MHD models, here the structure of the disturbance fields are again similar: both featuring overlapping alternating-sign vorticity structures adjacent to the side-wall, and a series of opposite-signed vertically aligned vortices adjacent to the near-wall vorticity structures. Some differences are observed: the outer regions of the side-wall layers display a qualitative discrepancy between models in the creation of cylindrical field-aligned vortices. For the full-MHD model, a quadratic three-dimensionality persists in the vortices to resemble the barrel-like turbulent structures theorised in \citet{Potherat:00} and reported in \citet{Muck:00} and \citet{Potherat:12}. In addition, the difference in structures across the two sub-layers within the Shercliff layer are reminiscent of that found in turbulent MHD flows by \citet{Krasnov:12}, albeit with slightly different topology dimensionality. Nonetheless, the closeness of eigenvalue predictions highlights an interesting facet of the SM82 formulation.  Although the exact nature of the flow may present small but noticeable differences, the averaging of flow dynamics along the field direction results in the mean integrated quantities being suitably approximated. As evidenced in figure~\ref{topt_q2d_v_3d}, this is even more-so for higher field strengths, where at low $R_S$ (applicable to fusion applications) the difference in topology is suggested here to be negligible.

\subsection{Transient growth mechanisms}
\label{sec:TG_mec}

This section quantifies the mechanisms leading to transient amplifications in an attempt to address whether instability in these duct flows can result exclusively from two-dimensional dynamics. An analysis of the energy growth in orthogonal velocity components is provided in \S~\ref{comp_en_sec}. To understand the fundamental flow dynamics which modulate mode topology and alter attainable energy gains, an analysis of the physical time-scales for inertial and electro-magnetic phenomena is given in \S~\ref{time_scal_sec}. However, before proceeding with this analysis, the fundamental mechanisms which lead to transient growth wall-bounded flows are revisited. 

For low-$\Ha$ MHD and hydrodynamic flows, transient amplifications can exist due to the growth of vertically-aligned and streamwise-aligned modes, which are respectively analogous to the well known Orr--Sommerfeld and Squire modes arising from analysis of the linear stability of parallel shear flows \citep{Orr:07,Landahl:80}. Each of these have their own fundamental mechanisms for transient energy growth. The two-dimensional (here magnetic field-aligned) Orr--Sommerfeld modes undergo modest transient amplification due to the Orr-mechanism; the wall-normal velocity disturbance leans upstream against the shear flow and raises up into Tollmien--Schlichting wave-packets which are subsequently propagated downstream \citep{Orr:07}. In contrast, the streamwise invariant Squire modes achieve significant transient amplification via the lift-up mechanism; low streamwise velocity fluid closer to the duct wall is \textit{lifted up} by the streamwise vortices, with higher velocity fluid subsequently replacing it \citep{Landahl:80}. Beyond these fundamental mechanisms, a coupling exists between modes in which the transient growth is further amplified by the forcing of streamwise-aligned vorticity by wall-normal velocity \citep{Schmid:01}.

\subsubsection{Componental energy analysis}
\label{comp_en_sec}
\begin{figure}
\begin{subfigure}[t]{0.45\textwidth}
\small (\textit{a})
\end{subfigure}\hfill
\begin{subfigure}[t]{0.48\textwidth}
\small (\textit{b})
\end{subfigure}\\
\centering
\begin{subfigure}[t]{0.5\textwidth}
\includegraphics[width=\linewidth,valign=t]{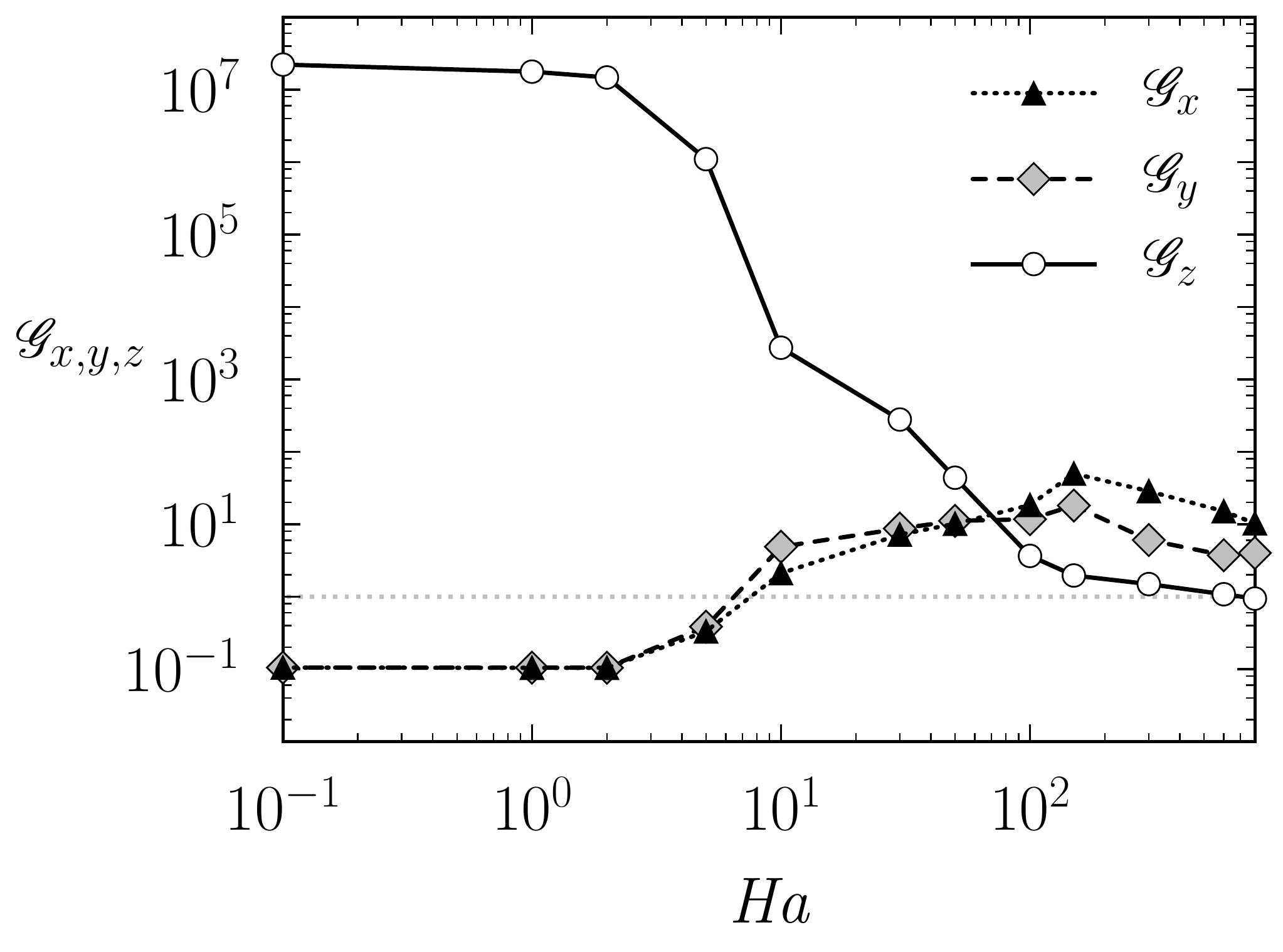}
\end{subfigure}\hfill
\begin{subfigure}[t]{0.5\textwidth}
\includegraphics[width=\linewidth,valign=t]{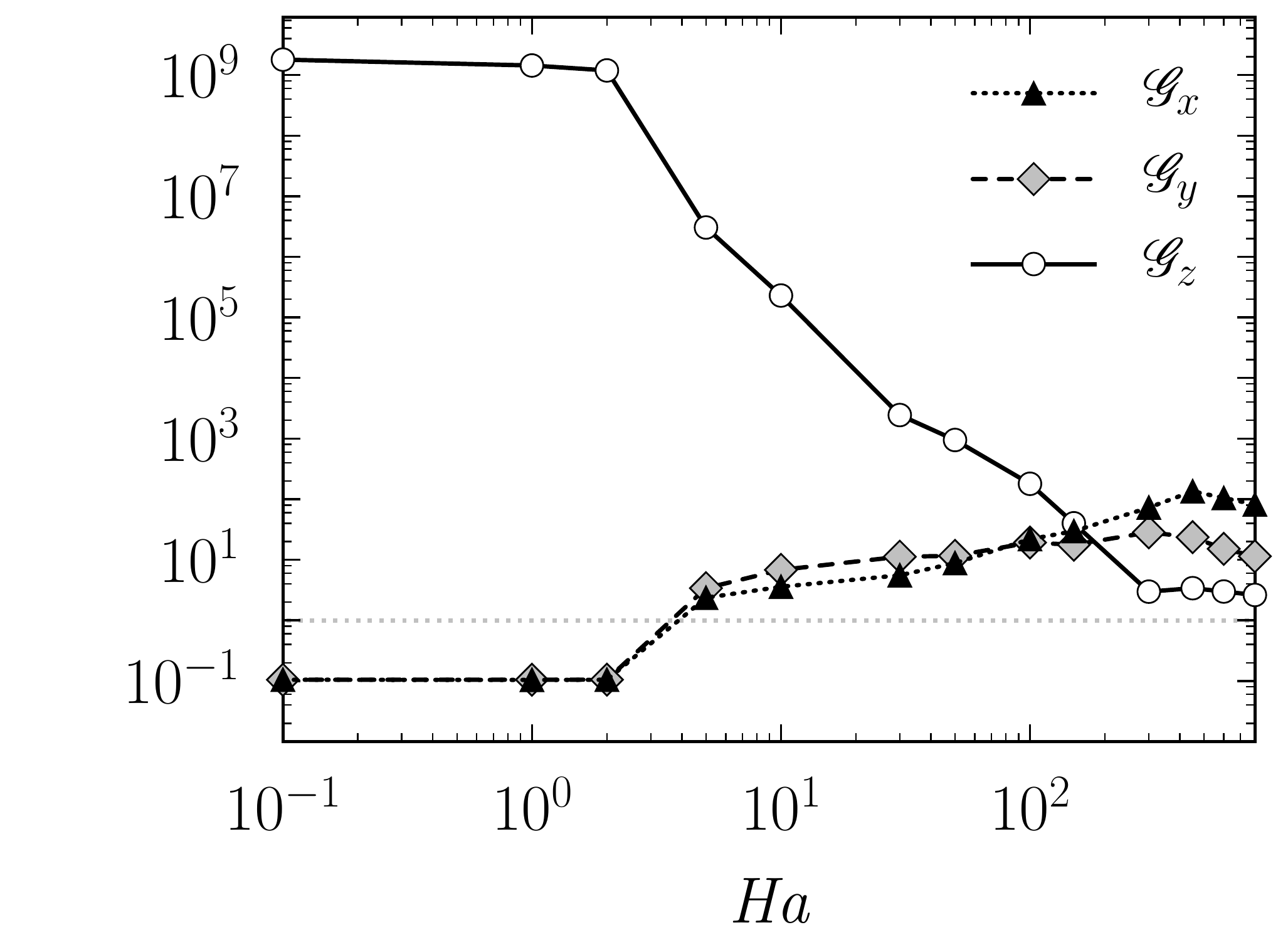}
\end{subfigure}\\
\caption{Amplification of perturbation kinetic energy $\mathscr{G}_x$, $\mathscr{G}_y$ and $\mathscr{G}_z$ in the horizontal, vertical and streamwise velocity components (i.e. $u,v$ and $w$), respectively, as a function of $\Ha$ for (\textit{a}) $\Rey =5000$ and  (\textit{b}) $\Rey =15 \,000$.}
\label{fig:vel_eigv}
\end{figure}

To characterise the dominance and coupling of specific transient growth mechanisms at varying $\Ha$ over evolution to $t=\TO$, the componental kinetic energy growth for the three-dimensional optimal disturbance is analysed. The energy gain in the horizontal, vertical and streamwise velocity components, represented respectively by $\mathscr{G}_x$, $\mathscr{G}_y$ and $\mathscr{G}_z$, is shown as a function of $\Ha$ for $\Rey =5000$ and $15\,000$ in figures~\ref{fig:vel_eigv}(\textit{a},\textit{b}), respectively.
%
%\BEQ
%\label{eq:comp_en}
%\mathscr{G}_i  \equiv  \frac{\int_\mathcal{\Omega}u_i^2(t=\tau_\text{opt}) \,\rm{d} \Omega}{\int_\mathcal{\Omega}u_i^2(t=0) \,\rm{d} \Omega},\qquad i = 1, 2, 3
%\EEQ
%where the subscript $i$ refers to the $i$th velocity component (i.e. $(u_1,u_2,u_3)\equiv (u,v,w)$). 

For low field strengths, transient growth is confined solely to the streamwise velocity components $\mathscr{G}_z$. This is strongly indicative of lift-up mechanisms dominating transient amplifications, and is consistent with the presiding hydrodynamic response. However, for $\Ha > 1$, a significant reduction in $\mathscr{G}_z$ correlated with an increase in spanwise components $\mathscr{G}_x$ and $\mathscr{G}_y$ signals impediment of this mechanism and a growing influence of the Orr-mechanism forcing sidewall-normal velocity. Regardless, the largest transient gains remain driven by lift-up responses, and only for $500 \gtrsim R_H \gtrsim 33.\dot{3}$ and $3000 \gtrsim R_H \gtrsim 50$ is significant growth found in all three velocity components at $\Rey = 5000$ and $\Rey=15\,000$, respectively. At transition to the Q2D-MHD regime, Orr related mechanisms become the largest component of transient growth for both Reynold numbers. Hence, for $R_H \lesssim 33.\dot{3}$ the lift-up mechanisms are largely suppressed, and the dominant two-dimensional Orr-mechanisms can be sufficiently captured by the SM82 formulation.  Transient growth is produced by the non-modal interaction of linear eigenmodes: revisiting figure \ref{Gmax}, the substantially higher optimal growths found in the three-dimensional duct at smaller $\Ha$ compared to the SM82 predictions can be understood in terms of the SM82 model having only a subset (the Orr modes) of the eigenmodes available to the three-dimensional model to interact for transient growth.

\begin{figure}
\centering
\begin{tabular}{l}
\\ \small (\textit{a}) $\Ha=10$, $R_H=500$, $R_S=1581.14$ \hskip 35pt \small (\textit{b}) $\Ha=50$, $R_H=100$, $R_S=707.11$\\
\begin{subfigure}{.5\textwidth}
  \centering
  \includegraphics[width=1\linewidth]{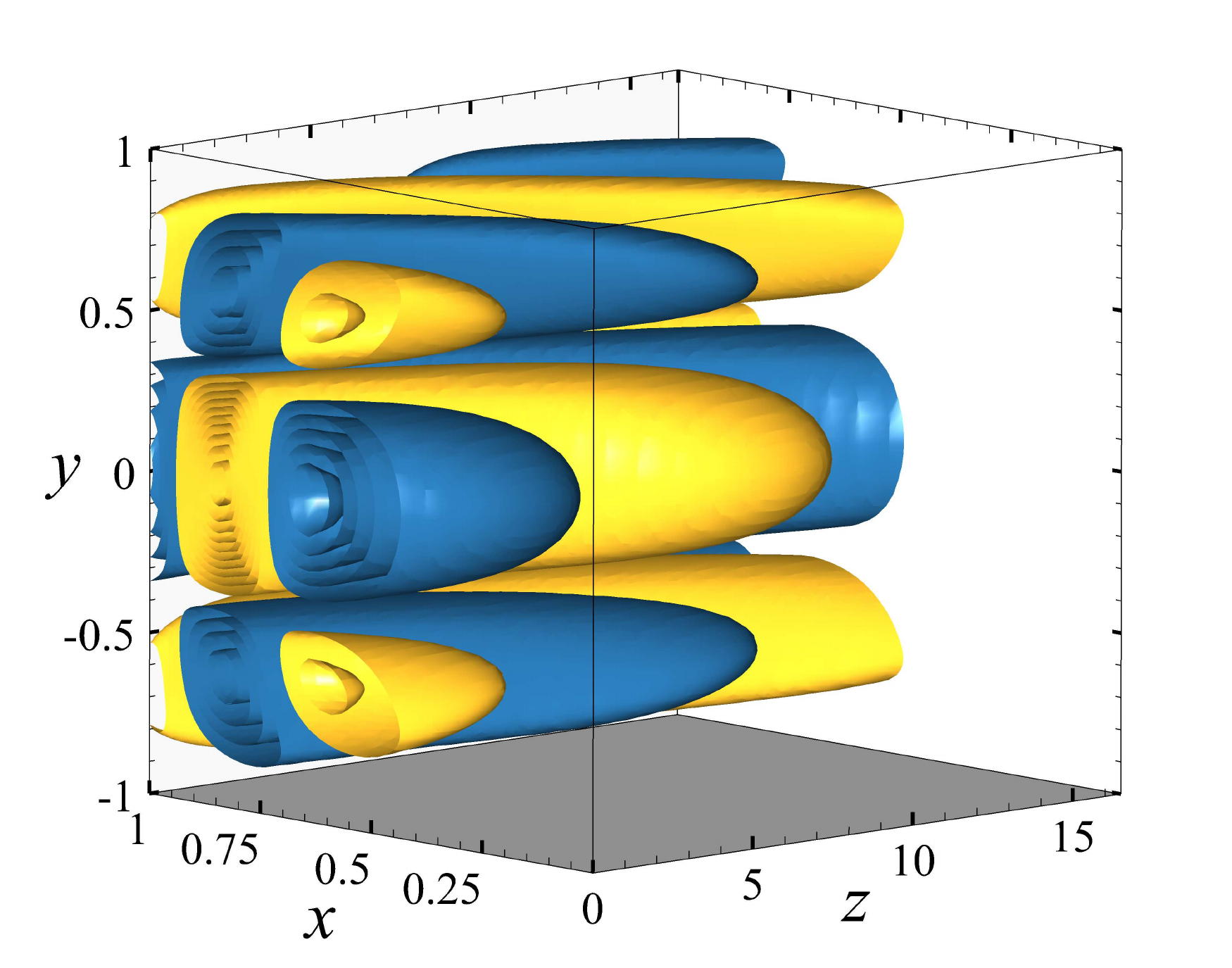}
\end{subfigure}
\begin{subfigure}{.5\textwidth}
  \centering
  \includegraphics[width=1\linewidth]{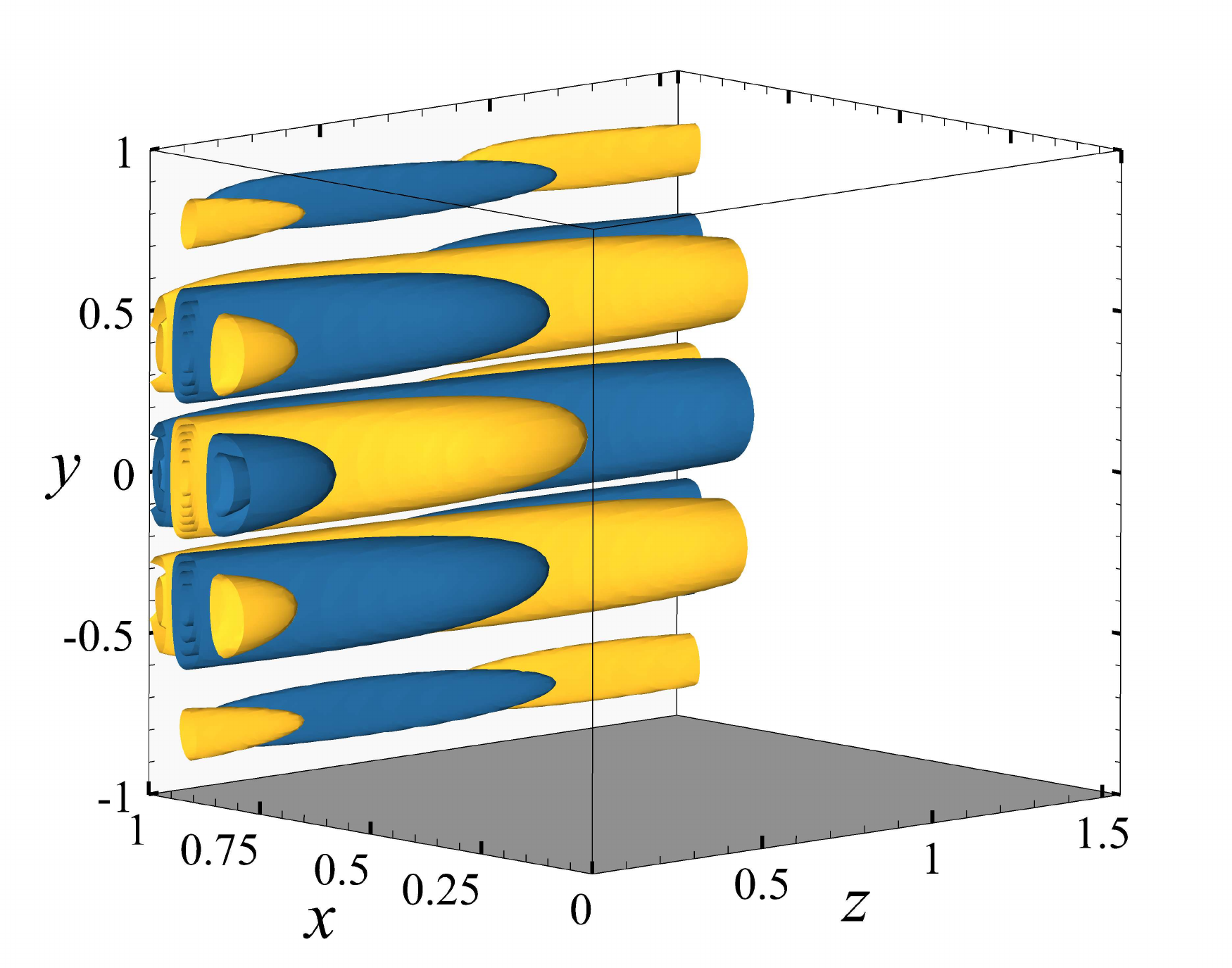}
\end{subfigure}
\\\\ \small (\textit{c}) $\Ha=150$, $R_H=33.\dot{3}$, $R_S=408.25$ \hskip 35pt \small (\textit{d}) $\Ha=300$, $R_H=16.\dot{6}$, $R_S=288.68$\\
\begin{subfigure}{.5\textwidth}
  \centering
 \includegraphics[width=1\linewidth]{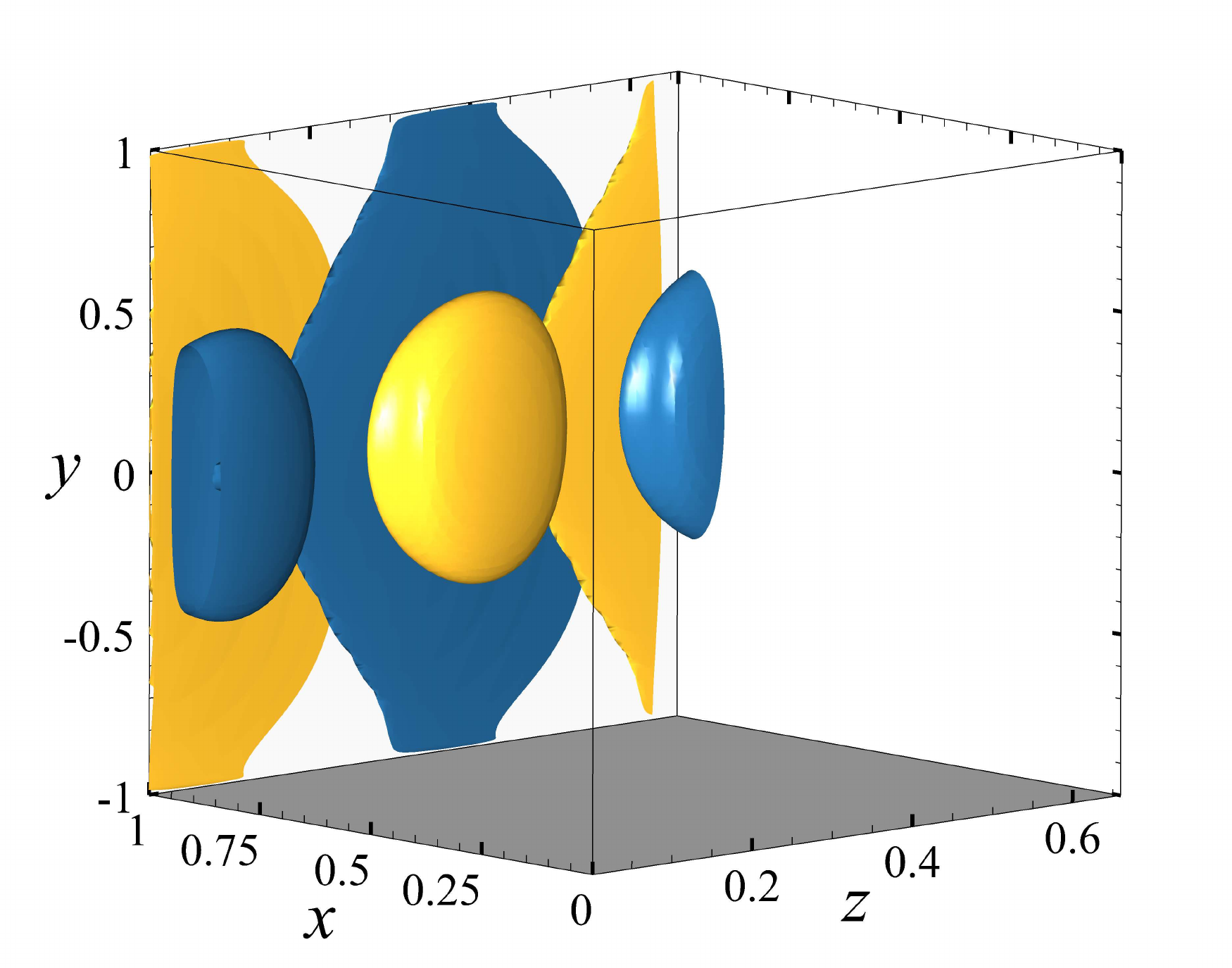}
\end{subfigure}
\begin{subfigure}{0.5\textwidth}
  \centering
 \includegraphics[width=1\linewidth]{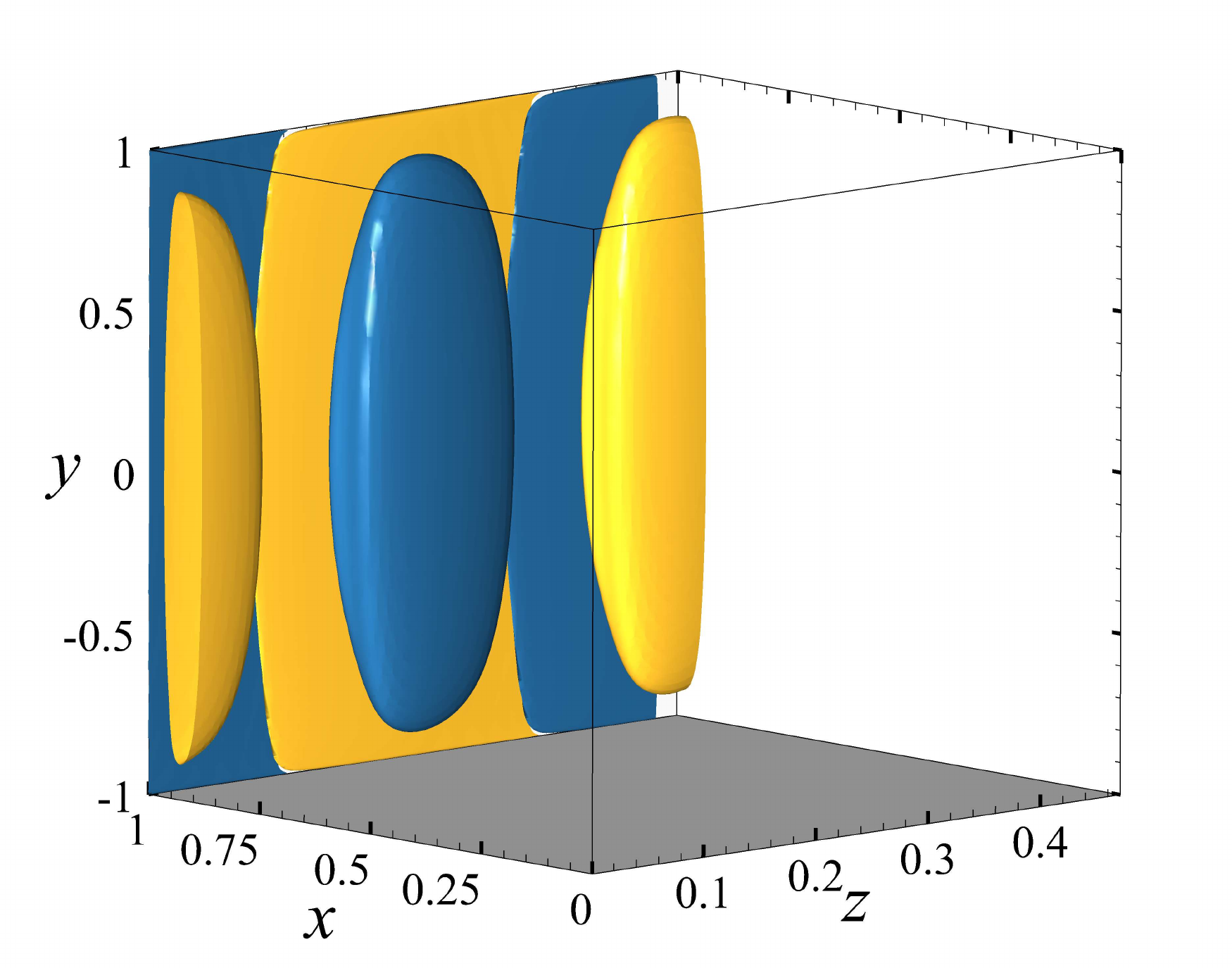}
\end{subfigure}
\end{tabular}
\caption{Vertical component of vorticity $\omega_{y}$ iso-surfaces of the perturbation fields at time of maximum transient amplification $\TO$ for (\textit{a}--\textit{d}) respectively $\Ha=10$, $50$, $150$ and $300$ at $\Rey=5000$. Iso-surfaces specified as per figure~\protect\ref{eig_xz_vort_Re5k}.}
\label{topt_eigv}
\end{figure}

To provide a qualitative picture of these effects, the energy contained in orthogonal velocity components is qualitatively captured through comparison of the optimal eigenvectors for $\Rey =5000$ in figures~\ref{eig_xz_vort_Re5k}(\textit{c}-\textit{f}), and their respective topology after time evolution to $\TO$ as seen in figure~\ref{topt_eigv}. For moderate-MHD regimes (figures~\ref{topt_eigv}\textit{a},\textit{b}), the perturbation field persists as streamwise aligned vortices consistent with lift-up mechanisms, yet now with their orientation slanted down-stream due to bulk shear caused by a mild Orr related response. For higher field strengths pertaining to the Q2D-MHD regime (figures~\ref{topt_eigv}\textit{c},\textit{d}), modulation of the optimal modes is again split into two regions inside the Shercliff boundary layers with varying degrees of two-dimensionality. However, the dominant disturbance structures remain aligned with the magnetic field, and their mean-radial growth is strongly reminiscent of Orr-mechanisms providing moderate transient growth. Here suppression of the lift-up response through the lack of any streamwise orientation is qualitatively seen. The correlation of this behaviour with the preclusion of streamwise energy growth $\mathscr{G}_z$ further explains why the SM82 model is an excellent predictor of transient growth mechanisms towards large $\Ha$.

\subsubsection{Time-scale analysis}
\label{time_scal_sec}
This section provides an analysis of the specific role electromagnetic and inertial phenomena play in determining transient growth and mode topology. To this end, the following non-dimensional time-scale definitions are introduced:
\vspace{0.5em}
\settasks{
	counter-format=(tsk[r]),
	label-width=4ex
	}
	
\begin{tasks}(1)
	\task Vertical mean-flow energy transfer
		\BEQ
		\label{mean_flow_ts_z}
		\tau_z = \frac{\ell_z}{U_0},
		\EEQ
		where $\ell_z$ is the streamwise wavelength of the optimal modes.
	\task Streamwise mean-flow energy transfer
		\BEQ
		\label{mean_flow_ts_y}
		\tau_y = \frac{\ell_y}{U_0},
		\EEQ
		where $\ell_y$ is the maximum vertical radius of a single vortex present in the optimal mode $\omega_y$ iso-surface,
	\task Viscous dissipation
		\BEQ
		\label{visc_ts}
		\tau_\nu = \Rey\left[\left(\ell_x^2\right)^{-1} + \left(\ell_y^2\right)^{-1} + \left(\ell_z^2\right)^{-1}\right]^{-1},
		\EEQ
		where $\ell_x$ is a characteristic length taken orthogonal to both $\boldsymbol{e}_y$ and $\boldsymbol{u}_0$, here, taken to be the Shercliff layer thickness $ \delta_{S}$,	
	\task Two-dimensionalisation by the Lorentz force
		\BEQ
		\label{lorentz_ts}
		\tau_L = \frac{1}{N}\left[1+\left(\frac{\ell_y}{\ell_x}\right)^2 + \left(\frac{\ell_y}{\ell_z}\right)^2\right],
		\EEQ
	\task Hartmann friction 
		\BEQ
		\label{hart_ts}
		\tau_H = R_H,
		\EEQ
	\task Total bulk dissipation 
		\BEQ
		\label{tot_ts}
		\tau_d = \left[\tau_\nu^{-1} + \tau_L^{-1}\right]^{-1},
		\EEQ
\end{tasks}

All time-scale quantities here are calculated at the optimal time-period $\TO$. All equations \eqref{mean_flow_ts_z}--\eqref{tot_ts} are considered in non-dimensional form with dimensional scales consistent with those defined in \S~\ref{prob_form}.

\begin{figure}
\begin{subfigure}[t]{0.45\textwidth}
\small (\textit{a})
\end{subfigure}\hfill
\begin{subfigure}[t]{0.48\textwidth}
\small (\textit{b})
\end{subfigure}\\
\centering
\begin{subfigure}[t]{0.5\textwidth}
\includegraphics[width=\linewidth,valign=t]{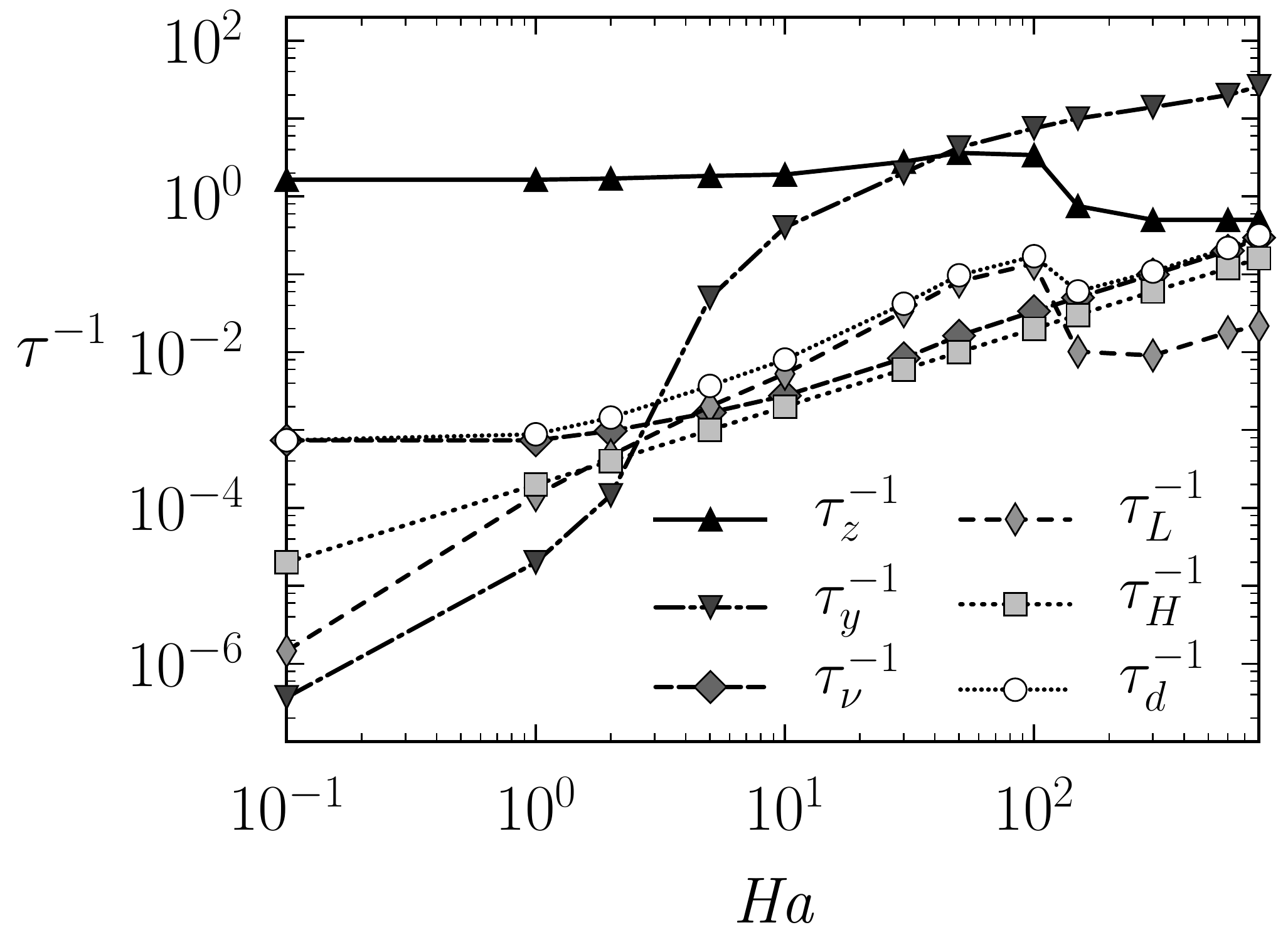}
\end{subfigure}\hfill
\begin{subfigure}[t]{0.5\textwidth}
\includegraphics[width=\linewidth,valign=t]{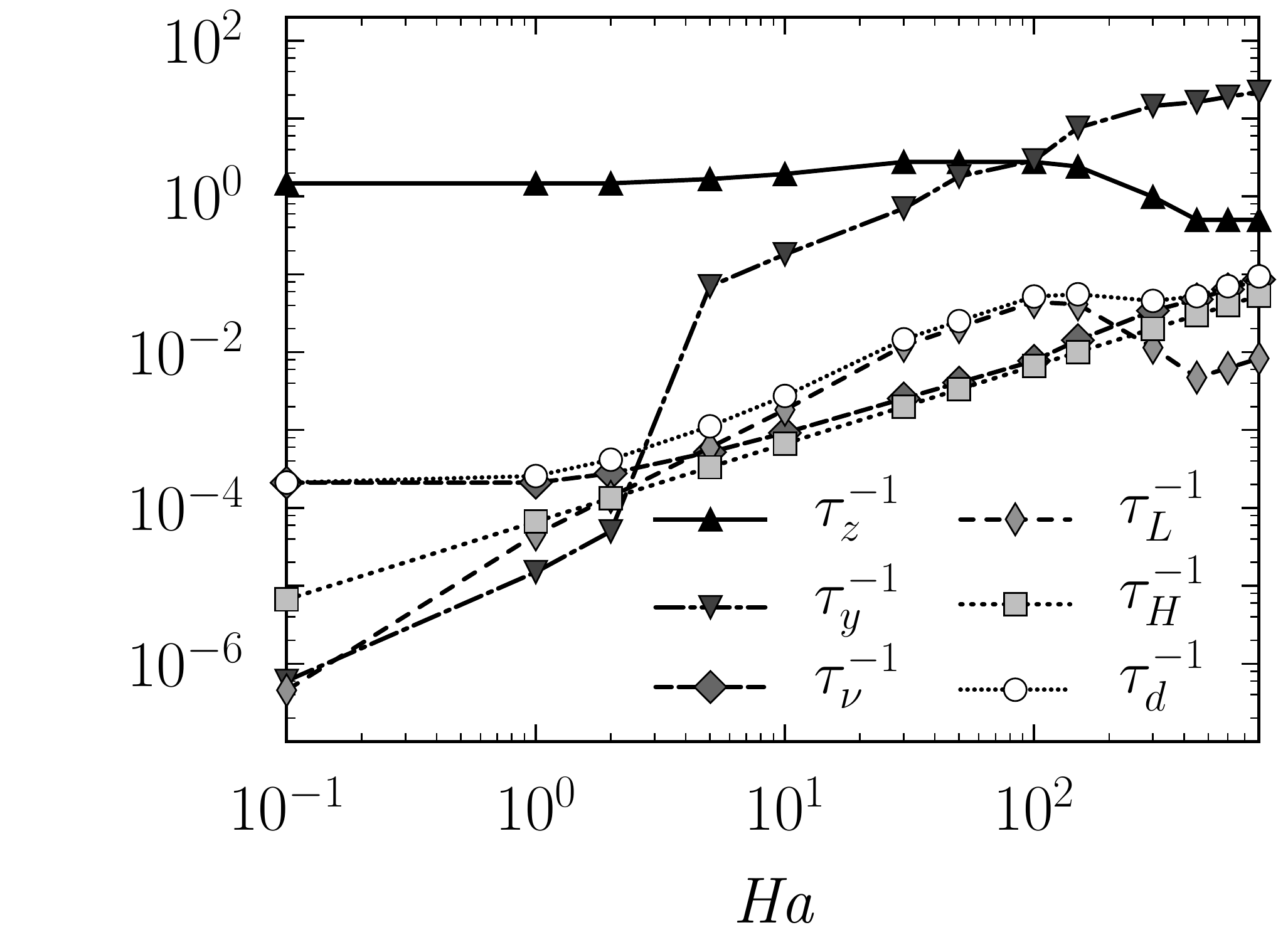}
\end{subfigure}\\

\caption{Inverse time-scales for $\Rey=5000$ and $15\,000$ are respectively plotted as a function of $\Ha$ in (\textit{a}) and (\textit{b}).}
\label{fig:tim_sca}
\end{figure}

Equations \eqref{mean_flow_ts_z}--\eqref{tot_ts} largely fall into one of two categories; those relating purely to inertial phenomena, and those not influenced by electromagnetic forces. The former consists of the time-scale for significant transference of energy from the mean baseflow to the perturbation field in the vertical and streamwise directions $\tau_z$ and $\tau_y$, respectively (analogous to the eddy turnover time, \citet{Frisch:95}), in addition to the time-scale for the dissipative effects of viscosity to act on the perturbation fields $\tau_\nu$ . The latter category comprises the time scale for the Lorentz force to suppress velocity differentials between transverse planes $\tau_L$, as well as that of energy dissipation due to friction in the Hartmann boundary layers $\tau_H$ (derived from the SM82 model, \citet{Potherat:00}). The total dissipative effects in the bulk flow due to both viscous and Lorentz forces are characterised by $\tau_d$. The inverse of each time scale defined in \eqref{mean_flow_ts_z}--\eqref{tot_ts} is presented for $0.1\leq\Ha\leq 800$ at $\Rey=5000$ and $15\,000$ in figures \ref{fig:tim_sca}(\textit{a},\textit{b}), respectively. Larger magnitudes of $\tau^{-1}$ indicate a given physical mechanism modulating the dynamics of the linearised flow at a faster rate. 

For all regime parameters investigated, the fastest acting mechanisms are one of the two inertial transfer times scales. For low- and moderate-MHD regimes, the lift-up related $\tau_z$ acts most quickly, whereas the transition to the Q2D-MHD regime sees a switch to the Orr-mechanism related streamwise equivalent $\tau_y$. This highlights an interesting point, despite the Lorentz force acting to bias the eigenspectrum towards Q2D eigenmodes with increasing Hartmann number, the growth mechanisms stem largely from inertial effects. In line with that presented for hydrodynamic flows by \citet{Barkley:08}, this suggests that the non-orthogonality driving transient growth emanate from the asymmetry of the advection operator describing inertial forces, rather than those relating to pressure (and here, Lorentz forces). For $Ha\gtrsim 5$, the hierarchy of time-scales changes in two significant ways. Firstly, Joule dissipation now acts more quickly than viscous dissipation. Secondly, the vertical inertial transfer time becomes faster than all other dissipative mechanisms at play. This change marks the transition to a regime where the transient growth is dominated by electromagnetic effects in the bulk. From this point onwards, the maximum amplification results from both inertial transfer terms, which are now also predominantly impeded by Joule dissipation. 

Importantly, at $R_H \approx 33.\dot{3}$ the Joule dissipation in the bulk suddenly falls below the level of the dissipation in the Hartmann layer derived from the SM82 model. This signals a transition to two-dimensional dynamics where bulk velocity gradients along the magnetic field have been smoothed out by the action of the Lorentz force, and the dissipation concentrates where the strongest gradients remain --- in the Hartmann layer. It is noteworthy that significant gradients still exist in the bulk of the optimal modes at the crossover point between the two regimes. Nevertheless, the transition also corresponds to the point where transient growth becomes controlled by $R_S$, which expresses the ratio between two-dimensional inertia and the Lorentz force, rather than $R_H$ in the three-dimensional MHD regime. Thus, although the point where bulk dissipation drops below the level of Hartmann layer dissipation is an unmistakable signature of a transition to quasi-two-dimensional MHD dynamics, the optimal mode retains noticeable three-dimensional features at values of $\Ha$ well beyond this transition \citep{Potherat:14}. Though somewhat counter-intuitive, the occurrence of two-dimensional dynamics has been recently observed in three-dimensional MHD turbulence by \citet{Baker:18}, and underlines that the link between topological and dynamical dimensionality is anything but obvious.

\section{Conclusion}
\label{conc}

Utilising a high order spectral element method in combination with a unique high order temporal scheme, the exact nature of optimal transient growth in MHD duct flow was elucidated for an extensive set of flow regime parameters. Beyond understanding the transitional nature of transient growth from three-dimensional to Q2D regimes, this work addressed the important question of whether purely Q2D transient growth mechanisms exist at adequately high and realistic Hartmann numbers, and if so, whether they can be accurately captured by means of the computationally affordable SM82 model. The answer to which is seminal to understanding the fundamental subcritical turbulent transition scenarios in MHD flow. The present analysis strongly suggests that the formation of Q2D turbulence may indeed be driven by Q2D mechanisms towards high $\Ha$, and the SM82 model was shown to be an excellent predictor of such dynamics.

Through a comparison of both 3D and Q2D MHD models, the optimal growth residing from the two approaches showed striking agreement in terms of both scaling $\Ha^{-1/2}$ and also magnitude, for sufficiently high field strengths $R_H \gtrsim 33.\dot{3}$. Despite slight eigenvector three-dimensionality persisting in the full MHD analysis, the mean topological characteristics were suitably approximated by the SM82 model upon transition to this regime. Furthermore, the results suggest that any discrepancies in mode topology was expected to become negligible at field strengths applicable to fusion reactor designs.

In addition to these properties, an essential measure of SM82 accuracy, that of the ability to resolve fundamental growth mechanisms, was also shown to be excellently captured using this numerical approach. Upon transition to the high-$\Ha$ regime, modest growth was due solely to the two-dimensional Orr-mechanism evolving two-dimensional Orr--Sommerfeld modes. Three-dimensional lift-up dynamics are precluded in this regime, and as such, the SM82 excellently resolves these fundamental growth attributes. Further support for two-dimensionalistation of the linear flow dynamics was also found in the switch of dominance from vertical to streamwise inertial time scales, in combination with a characteristic drop of Joule dissipation below that of Hartmann friction.

This validation of the SM82 model's effectiveness in resolving key linear growth properties such as energy gain, mode topology and growth mechanisms, allows for three-dimensional numerical constraints to be circumvented, which in turn opens the door for a full description of the transition to turbulence at high $\Ha$. Future research would be well served to investigate the stability of the quasi-two-dimensional MHD solutions in industry-applicable high-$\Ha$ regimes. The purpose being to ascertain the role played by the Orr--Sommerfeld modes in the transition at Reynolds numbers in the vicinity of criticality; potentially having a meaningful impact on meeting nuclear fusion reactor viability constraints.

A final few words of caution on the validity of the SM82 formulation with respect to magnetic confinement nuclear fusion reactor design are in order. The SM82 model has proven its validity on simple geometry such as ducts. However, inertial effects at locations of spatial discontinuities (i.e. sharp corners) could still incur three-dimensionality, even at very large field strengths. Nevertheless, as this has not been presently investigated in any depth, the three-dimensional analysis may still end up validating the SM82 model in these cases; just as it has been shown for the case of a wakes behind obstacles \citep{Dousset:08, Kanaris:13}.

\section{Acknowledgements}
O.G.W.C.\ was supported by an Engineering Research Living Allowance (ERLA) scholarship from the Faculty of Engineering, Monash University. This research was supported by Discovery Grants DP150102920 and DP180102647 from the Australian Research Council, and was undertaken with the assistance of resources from the National Computational Infrastructure (NCI), which is supported by the Australian Government. This research was also supported by the Royal Society International Exchanges Grant (Grant No. IE170034). A.P.\ acknowledges support from the Royal Society under the Wolfson Research Merit Award Scheme (Grant No. WM140032).

%\bibliographystyle{jfm}
%\bibliography{../../Library/master/MHD}

\end{document}